\documentclass[12pt]{article}

\pdfoutput=1 

\usepackage[left=1.5cm, right=1.5cm, top=1.5cm, bottom=1.5cm]{geometry}

\usepackage{graphicx}
\usepackage{eqnarray}
\usepackage{cite}
\usepackage{multicol}
\usepackage{caption}
\usepackage{subfigure}
\usepackage{lscape}
\usepackage{amssymb}
\usepackage{breqn}

\newcommand\impliess{\Rightarrow}

\newcommand{\keywrds}{\textbf{Keywords: }}
\newcommand{\pacs}{\textbf{PACS: }}
\Large
\title{Accretion onto a Quintessence Contaminated Rotating Black Hole : Violating the Lower Limit for Eta over s}

\date{}

\author{Ritabrata Biswas $^1$ $\;\;\;\;\;\;\;\;$Promila Biswas $^3$ $\;\;\;\;\;\;\;\;$ Parthajit Roy  $^{2*}$\\ \\$^{1,3}$ Department of Mathematics, The University of Burdwan\\Golapbag Academic Complex, City:Burdwan-713104,\\Dist: Purba Bardhaman, State: West Bengal, India\\$^1$e-mail: \texttt{biswas.ritabrata@gmail.com}  \\
$^3$e-mail: \texttt{promilabiswas8@gmail.com}  \\
\\$^2$Department of Computer Science,\\ The University of Burdwan\\City:Burdwan-713104,\\Dist: Purba Bardhaman, State: West Bengal, India\\e-mail: \texttt{roy.parthajit@gmail.com} \footnote{$^2$ \emph{ Corresponding Author:}Parthajit Roy, roy.parthajit@gmail.com}}

\begin{document}
	\maketitle
	
	\begin{abstract}
	Viscous accretion flow around a rotating supermassive black hole sitting in a quintessence tub is studied in this article. To introduce such a dark energy contaminated black hole's gravitational force, a new pseudo-Newtonian potential is used. This pseudo-Newtonian force can be calculated if we know the distance from the black hole's center, spin of the black hole and equation of state of the quintessence inside which the black hole is considered to lie. This force helps us to avoid complicated nonlinearity of general relativistic field equations.  Transonic, viscous, continuous and Keplerian flow is assumed to take place. Fluid speed, sonic speed profile and specific angular momentum to Keplerian angular momentum ratio are found out for different values of spin parameter and quintessence parameter. Density variation is built and tallied with observations. Shear viscosity to entropy density ratio is constructed for our model and a comparison with theoretical lower limit is done.
	\end{abstract}

\keywrds{Accretion Disc, Viscosity, Quintessence, Supermassive Black Holes, Shear Viscosity to entropy ratio}


\section{Introduction}
\label{sec:introduction}

Since 1998, despite the propositions of different conceptual models, we were not really able to construct a hundred percent correct theoretical support to the observations of late time cosmic acceleration. Stress energy tensor was redesigned to interpret the accelerated expansion. The names of candidates of such a repulsive energy/ hypothetical matter are familiar as dark energy (DE)~\cite{1:garnavich:1998}.  $\Lambda $CDM (Cosmological constant with cold dark matter) model is successful to describe the formation and the evolution of Large Scale Structures in the universe~\cite{1:new:Spergel:2003,6:new:Smith:2007}. However, problems like fine tuning~\cite{9:new:RevModPhys} and cosmic coincidence are faced~\cite{10:new:Astashenok:2012}. We will discuss in detail the anomalies faced by $\Lambda $CDM model. Firstly, the lithium abundance challenge should be discussed. It is followed~\cite{1982:Spite} that no significant destruction of $Li^7$ took place during the protostellar phase of solar mass stars. It is assumed that the lithium abundances of old halo stars is representative of abundance in the primordial matter. This does not signify $\Lambda $CDM. Next anomalies to be discussed must be CMB power asymmetry, missing satellites, very massive galaxies at high redshift etc. $\Lambda $CDM model is also found to encounter problems while describing structures at small scales\cite{31:moore:1994,32:Moore:1999,33:Ostriker:2003,34:Boylan:Kolchin:2011,35:Boylan:Kolchin:2012,36:Oh:2011}. Among these, main problems are CUSP/ CORE problem   \cite{31:moore:1994,37:Del:Popolo:2017,37:new:Flores:1994}, saying about the flat density profiles of dwarf galaxies, irregulars and low surface brightness galaxies. Missing satellite problem coins the anomaly between the number predicted subhalos in N-body Simulations~\cite{32:Moore:1999,41:new:Klypin:1999} and those actually observed. This is further complicated by the “to big to fail” problem, arising from the $\Lambda $CDM prediction of satellites that are too massive and dense, compared to those observed~\cite{34:Boylan:Kolchin:2011,35:Boylan:Kolchin:2012}. Similarly, we face the problems regarding angular momentum catastrophe~\cite{42:new:van:den:Bosch:2001,43:new:Cardone:2009} and alignment on thin planes of satellite galaxies of MW and M31 is difficult to explain in simulations of the $\Lambda $CDM paradigm~\cite{44:new:Pawlowski:2014}. The problem of obtaining the slope and scatter of baryonic Tully-Fisher relation ($M_b \propto V_c^4$)~\cite{45:new:McGaugh:2011}. Other issues are discussed in references~\cite{46:new:Kroupa:2005,47:new:Kroupa:2010,48:Kroupa:2012,49:new:KROUPA:2012,50:new:Kroupa:2015}. Lastly, we recall about the cosmic coincidence problem~\cite{12:new:Sivanandam:2013} which can not be justified from the $\Lambda $CDM model.

It is noted that to support cosmic acceleration, matter energy part staying inside the cosmos must violate the strong energy condition $\sum_k\left(\rho_k+3p_k\right)<0$, where $\rho_k$ and $p_k$ denote energy density and pressure of different components filling up the cosmos. Standard approach leads up to consider DE a fluid exerting negative pressure and possessing barotropic equation of state $p=\omega \rho$, where $\omega$ is constant being equal to $-1$. But $p=-\rho$ implies no dynamics.

Recent observational reconstructions of DE EoS interpret $\omega$ to be dynamical and to cross phantom divide $\omega=-1$~\cite{2:6a:new:Zhao:2017,3:6a:Wang:2018}. $\omega$ is preferably treated to be a function of time scale factor or redshift. We face again another problem that dynamical DE, $\omega=\omega(a)$, combined with the assumption of DE as a perfect fluid arises some thermodynamic problems (eg., positiveness of entropy, temperature and chemical potential leads to $\omega \geq -1$ which contradicts the existence of phantom DE~\cite{4:6a:Duarte:2019,5:6a:Silva:2013}.)

To pass by such thermodynamic conflicts, we suppose DE to be a fluid with which bulk viscosity is present, i.e., a break for the perfect fluid hypothesis. Fluid with bulk viscosity supports dissipative processes. This again allows violation of dominant energy condition $p+\rho<0$ even without the DE necessarily becomes phantom~\cite{6:6a:BARROW1988743,7:6a:CRUZ2017103}. With this model, the late time cosmic  acceleration be explained as the effect of negative pressure due to bulk viscosity, $\zeta<0$, where $p_{eff}=p+\zeta$~\cite{8:6a:Zimdahl:2001,9:6a:Balakin:2003}. Eckart theory~\cite{10:6a:PhysRev} has been used in maximum cases to build such models to cross phantom line~\cite{11:6a:Brevik:2005}, magnitude of viscosity~\cite{12:6a:Brevik:2015}, consideration of Big Rip singularity~\cite{13:6a:Brevik:2013}, unified dark fluid cosmologies~\cite{14:6a:Avelino:2010,15:6a:Velten:2011} etc. We must refer two pioneer works for dissipative processes in cosmology~\cite{19:6a:Maartens:1995} and relativistic fluid dynamics and dissipative relativistic fluids along with their applications to cosmology and astrophysics and bulk viscous perturbations~\cite{20:6a:Maartens}. 

Cosmological perturbation theory along with non-ideal fluid in the presence of shear and bulk viscosities is constructed with the energy momentum tensor for non-ideal fluid~\cite{38:6a:Weinberg} given as

\begin{equation}
T^{\mu\nu}=\rho u^{\mu} u^{\nu} +\left(p+p_b\right)\Delta^{\mu\nu}+\Pi^{\mu\nu}~~~~,
\end{equation}

where $p_b(=-\zeta\nabla_{\mu}u^{\mu})$ is bulk viscous pressure with bulk viscosity coefficient $\zeta$. $\Pi^{\mu\nu}$ represents the shear viscosity tensor, having the form 

\begin{equation}
\Pi^{\mu\nu}=-2\eta \sigma^{\mu\nu}=-2\eta\left[ \frac{1}{2}\left(\Delta^{\mu\alpha} \nabla_{\alpha}u^{\nu}+\Delta^{\nu\alpha} \nabla_{\alpha}u^{\mu}\right)-\frac{1}{3}\Delta^{\mu\nu}\left(\nabla_{\alpha}u^{\alpha}\right)\right]~~~~,
\end{equation}

with $\eta$ being the shear viscosity. $\Delta^{\mu\nu}=u^{\mu}u^{\nu}+g^{\mu\nu}$, the projection operator, does project to the subspace orthogonal to the fluid velocity.

Nowadays, there are indirect ways to anticipate viscosity of the dark sector. But, before speaking about those indirect ways, we should focus on some references like~\cite{2:Turner:1984} \& \cite{3:mathews:2008} where  dark matter(DM)-DE interactive models are proposed to support cosmic acceleration and observed viscosity of DE. These references have considered the phenomena of  ``delayed decay of cold DM'' (from initial matter dominated flat cosmology with $\Lambda=0$) into light undetectable relativistic species. This generates bulk viscosity and a $\Lambda=-1$ epoch\footnote{actually measured as $-1.028\pm 0.032$ by the Planck Collaboration (2018) \cite{lewis:2020}. True dimension of $\Lambda$, however, becomes equivalent with $length^{-2}$. According to the reference\cite{lewis:2020}, dimensionless density parameter $\Omega_{\Lambda}=0.6889\pm 0.0056$, value of Hubble’s parameter at present time = $67.66\pm0.42~Kms^{-1}/Mpc=2.1927664\pm 0.0136\times 10^{-18} s^{-1}$ and the value of $\Lambda$ becomes $\Lambda=3\left(\frac{H_0}{c}\right)^2\Omega_{\Lambda}=1.1056\times10^{-52}m^{-2}=2.888\times\times10^{-122}l_p^{-2}~~,$ where $l_p$ is the Planck length.}
 near the neighborhood of present time. Deceleration for this model is found to be faster due to the presence of the matter content. Also the model is found to be consistent with the surveys of supernova magnitude-redshift relation~\cite{4:reiss:1998}\cite{5:perlmuttr:1998} and ages from the superannuated stars and globular clusters.

There exist different DE models. Quintessence and modified Chaplygin gas (MCG) are two important candidates among such models. The EoS of quintessence is given as 
\begin{equation}
p_q=\omega_q \rho_q.
\end{equation}
The value of $\omega_q$ regulates nature of the chronological evolution of universe. We may obtain the radiation, pressureless dust era, quintessence and phantom barriers for $\omega_q=\frac{1}{3}, 0, -\frac{1}{3}$ and $-1$ respectively. On the other hand, MCG has its EoS as 
\begin{equation}
p=\alpha \rho - \frac{\beta}{\rho^n} .
\end{equation} 

MCG can mimic different chronological stages of cosmic evolution depending on the value of $n$, mainly. Reference~\cite{71:new:Velasquez:Toribio:2011} uses type Ia supernovae and BAO data set to predict the best fit parametric values as $\alpha=0.061\pm 0.079$ and $n=0.053\pm 0.089$ (Best fit for Constitution+BAO+CMB) and $\alpha=0.110\pm0.097$ and $n=0.089\pm 0.099$ (Best fit for Union2+BAO+CMB). Another article~\cite{72:new:Lu:2010} uses Union2, SNIa, OHD, CBF, BAO and CMB data to constrain the modified Chaplygin gas model parameters with $1\sigma$ and $2\sigma$ confidences as $\alpha=0.00189^{+0.00583+0.00660}_{-0.00756-0.00915}$ and $n=0.1079^{+0.3397+0.4678}_{-0.2539-0.2911}$ .

Now, we will focus on these ``indirect ways'' to realize and measure the viscous properties of DE. To do so, we will take the help of gravitational wave (GW) signals. It is quite possible to constrain the constituents present in our universe by observing the nature of propagation of GW through them. Till date, GW signals from several black hole (BH) binary mergers like GW150914, LVT151012, GW51226, GW170104, GW170608, GW170814 etc and GW signals from neutron star binary like GW170817 etc are announced by LIGO and Virgo collaborations~\cite{6:abbott:2016}\cite{7:abbott:2016}\cite{8:abbott:2016}\cite{9:abbott:2017}\cite{10:abbott:2017}\cite{11:abbott:2017}\cite{12:abbott:2017}. Simultaneously, electromagnetic (EM) radiations coming out of the same sources have been detected. Tallying these, we are able to measure the arrival delay between the EM photons and GWs through the cosmological distances~\cite{13:will:1998}\cite{14:nishizawa:2016}\cite{15:li:2016}. Different references like~\cite{16:will:2014}\cite{17:khaya:2016} \& \cite{18:Wu:2016}  predict that GWs should propagate freely without any absorption and dissipation if perfect fluid medium is considered to be embedded in Friedmann-Robertson-Walker universe is considered. Nevertheless, this scenario changes as soon as the fluid content is chosen to be non-ideal type~\cite{19:hawking:1966}, GW gets dissipated with a damping rate $\beta_D\equiv 16\pi G_N\eta$~\cite{20:esposito:1971}\cite{21:madore:1973}\cite{22:prasanta:1999} when an amount of shear viscosity $\eta$ is incorporated into the fluid's energy momentum tensor, $G_N$ being the Newtonian gravitational constant. So, changes in $\beta_D$ which can be noticed in GW attenuation indicate changes in the value of $\eta$ over time or over cosmic distances. This is nothing but the evolution of viscosity, especially the shear viscosity, of DE over time. The authors of the reference~\cite{23:lu:2018} present a statistics of different GW events, median value of source luminosity distances with $90\%$ credible intervals in $Mpc$ units and upper limit on the damping rate $\beta_D$ at $95\%$ of confidence level in units of $10^{-3}~Mpc$ given by table~\ref{table:lumino}.

\begin{table}
	\caption{Luminosity Distances of different GW events}
	\centering
	\begin{tabular}{ |c|c|c| } 
		\hline
		GW Event & Luminosity Distance & $\beta$ \\  
		& ($MPc$)               & ($10^{-3} MPc^{-1}$) \\ \hline 
		LVT 151012 & $1000^{+500}_{-500}$ & 1.29 \\ \hline
		GW 170104 & $880^{+450}_{-390}$ & 1.40 \\ \hline
		GW 170814 & $540^{+130}_{-210}$ & 1.25 \\ \hline
		GW 151226 & $440^{+180}_{-190}$ & 2.39 \\ \hline
		GW 150914 & $410^{+160}_{-180}$ & 2.50 \\ \hline
		GW 170608 & $340^{+140}_{-140}$ & 3.05 \\ \hline
		GW 170817 & $40^{+8}_{-14}$ & 14.08 \\ 
		\hline
	\end{tabular}
	\label{table:lumino}
\end{table}

It is clear from these data that as we look through longer distances, i.e., we look in past, the shear viscosity reduces. We are able to conclude that DE, as time grows, exerts more and more amount of shear viscosity.

Regarding bulk viscosity of DE, especially for generalized Chaplygin gas type model, we can find several works. The reference~\cite{24:Li:2014} uses the then-available cosmic observational data from SNLS3, BAO, HST and Planck and constrains  the value of bulk viscosity coefficient as, $$\zeta_0= {{0.0000138^{+0.00000614} _{-0.0000105}}^{+0.0000145}_ {-0.0000138}}^{+0.0000212}_{-0.0000138}$$ However, even shear and bulk viscosity are related to each other~\cite{25:herzfeld:2004}.

Viscous effects of a fluid is more realizable when it flows; particularly layer by layer. Most prominent examples are the narrow X-ray binaries where an accretion disc is likely to get formed. The most simple diagram of such a Roche lobe overflow incident can be imagined with some assumptions : Consider that the flow is axis-symmetric, i.e., causing a cylindrical structure around a compact star-preferably a BH- where  $r$, $\phi$ and $z$ are the coordinates. We will consider $\frac{\partial}{\partial \phi}\equiv 0$ and also a stationary disc, i.e.,  $\frac{\partial}{\partial t}\equiv 0$. We will further consider a thin disc, i.e., $\frac{h(r)}{r}<<0$, where $h(r)$ is the disc height. The viscous effect of the disc is also taken to be small, i.e., the radial inward velocity $v_r<<$the local Keplerian rotational speed/azimuthal velocity $v_{\phi}=\Omega r=\frac{GM}{r}$, if the radial momentum is conserved. On the other hand, conservation of angular momentum requires effects of viscous forces to take into account. Keplerian balance always implies differential rotation. A transportation of angular momentum in the direction, perpendicular to the velocity, should be observed due to the differences in velocities at  different locations. If this is absent, translational (shear) viscosity will turn down to zero. To simplify the nonzero viscosity, a simplistic description of the physics of accretion disc can be obtained by $\alpha_{ss}$ \cite{26:shakura:1973}~prescription. Though the accretion driven/driving viscosity is of magnetic origin, it is popular to use an effective hydrodynamic description of the related disc presented by the hydrodynamic stress tensor as~\cite{27:landau:1987}:

\begin{equation}
\tau_{r\phi}=\rho\nu\frac{\partial v_{\phi}}{\partial r}=\rho\frac{d\Omega}{d\ln(r)}~~~~,
\end{equation}
where $\rho$ and $\nu$ are the density and kinematic viscosity coefficient respectively. Notifying total thermal pressure by $P$, isothermal sound speed $\sqrt{\frac{P}{\rho}}$ by $c_s$ and introducing a regulating parameter $\alpha_{ss} (\le 1)$, Shakura and  Sunyaev~\cite{26:shakura:1973}  proposed the prescription

\begin{equation}\tau_{r\Phi} = \alpha_{ss} P \impliess \gamma = \alpha_{ss} c^2_s\left[\frac{dr}{dh}\right]^{-1}.
\end{equation}
For Keplarian angular velocity $\Omega = \Omega_k = \left(\frac{G_NM}{r^3}\right)^{-1}$, we have,

\begin{equation}
\gamma = \frac{2}{3} \alpha_{ss} c^2_s/\Omega_k.
\end{equation}
To keep the equilibrium, the gravitational force is counter acted by the force produced by the pressure gradient

\begin{equation}
\frac{\partial p}{\partial z} = \rho , ~~~~ g_z = \rho \frac{G_{N}M}{R^2} \times \frac{z}{R}
\end{equation}
and as $h(\lambda) << r$, we obtain
\[
\frac{h}{r}\approx \frac{c_s}{v_k} \impliess \gamma \approx \frac{2}{3} \alpha_{ss} c_s h.
\]
This will be the way to replace the kinematic viscosity by $\alpha_{ss}$ parameter.

The value of $\alpha_{ss}$ is typically assumed to lie between 0.01 to 0.1~\cite{28:gou:2011}. Fromang et al.~\cite{29:fromang:2011} have found  a radially varying $\alpha_{ss}$, the overall size of which was over an order of magnitude lower, peaking at 0.013 and declining to below 0.002.

It is clear that, to measure viscosity, we will require to know the variation of density as well.  Again accretion density and many other properties are dependent on the spin parameter of the central gravitating BH. Fink~\cite{Fink:2016} found the spin parameter $a$ for Arakelian 120 galaxy in the constellation of Orion at coordinates $\alpha_{J2000.0}05^h 16^m 11.395^s \delta_{J2000.0} 00' 5^{9.65'}$ to be $a = 0.99 ^{+0.003}_{0.004}$. But Turner et. al.~\cite{2:Turner:1984}  shows the range to be $0.996\le a \le 0.998$. Again for the same Seyfert I galaxy, Fink measured the number density $n$ of the accretion disc to be $10^{15}cm^{-3}$. Reference~\cite{2:Turner:1984} has also measured it as $10^{15.95}cm^{-3}$. Mean molecular weight of Sun is chosen as $0.62$ (ionized gas) and hence the density of the taken accretion disc is found to be $\approx 6.2\times 10^{15} gm~cm^{-3}$.

Primary physical model of spherically symmetric gas accretion falling onto an astrophysical object was studied by Bondi for the first time. If rotation of the accreting fluid is not taken into account, effective accretion begins from a characteristic radius (So called Bondi radius, given by $R_B = \frac{GM}{c^2_s}$, $c_s$ being the sound speed through the gas.) by dominating the thermal energy by negative gravitational energy.  It has been considered that the density distribution to follow $\rho \propto \left(\frac{1+R_B}{r}\right)^{3/2}$.

Bondi and Shakura-Sunyaev considered accretion dynamics by considering Newtonian potential. The essential general relativistic effects on the curvature, i.e., the gravity around a BH was not taken into account. The later work has considered the only general relativistic effect by truncating the innermost edge of the disc at the last stable orbit of the Schwarzschild geometry. Novikov and Thorne~\cite{novikov:1973} have developed a complete general relativistic description of a thin Keplerian disc~\cite{ghosh:2007}.

To reduce general relativistic non-linearity, it is helpful to consider stationary flow and to replace  the general relativistic effect by the introduction of pseudo Newtonian potentials (PNP). Paczynsky \& Witta~\cite{paczynsky:1980} proposed such a force which exactly reproduces marginally stable orbit and marginally bound orbit of that infall GR. But this potential does consider only BH's mass. As almost all the celestial objects are rotating, the BHs are also rotating and Mukhopadhyay(2002)~\cite{BM2002:mukhopadhyay:2002} developed a PNP for a rotating BH for the first time. Sarkar and Biswas~\cite{sarkar:biswas:2019} has constructed a PNP for a rotating BH embedded in quintessence. This model proposes at most $4.95\%$ error as compared to  GR results. Roy and Biswas have modeled an accretion structure with Sarkar \& Biswas's potential. For this model, the first requisite cases should be known why accretion onto such a quintessence contaminated BH will take place. We will call the associated force as Pseudo Newtonian Force (PNF).

This is obvious to consider supermassive black holes (SMBHs) in the center of galaxies through the cases of such a presence is not justified. We are able to observe SMBHs at redshift $z=7.54$ which must have formed within less than one billion years~\cite{D1:deuardo:2018}. Alternative models to BHs, inclusion of extended objects in classical general relativity~\cite{D2:chirenti:2007}, consideration of existence of more exotic models, viz  ``naked singularity''~\cite{D3:joshi:2011} have been considered. So far, the motivations of these works were to consider only the gravitational effects of alternatives to BHs and to find out their observational properties in order to distinguish a BH from a so called BH mimicker. Recently, in the references like~\cite{D4:levkov:2018}, a possibility of DM, in the form of bosons, to form self gravitating bound structure in different galaxies are studied. Authors of~\cite{D5:boshkayev:2019} have compared the motion of test particles in the gravitational field of both SMBH and DM core. A significant discrepancy in the motion is noticeable around the radial distance $100AU$ and this increases as we are approaching to the center.

Finer observations in future (Say VLBI, BH cam project etc) might be able to distinguish the shadows caused by BH and BH mimicker. As of now, we cannot exclude the idea of existence of SMBH candidates like gravastars or boson stars etc. These studies/realizations motivate us to consider quintessence contaminated BHs. Besides, DM clustering are chosen to be the cause of formation of different structures of universe, especially the galaxies. DM and DE interact. As mentioned earlier, DE and bulk viscosity even can be formed out of the delayed DM decay. As a result, we can expect the presence of DE at the vicinity of the core area of SMBHs. This motivates us to study the viscous accretion onto quantum contaminated SMBHs.

Another motivation for the present work must be mentioned here. While studying the accretion and wind properties, we see for adiabatic fluid, wind branches are almost parallel to $x$ axis in $u-x$ plane while we go far from the central BH. On the contrary, the wind branch turns to be parallel to $u$ axis while modified Chaplygin gas is accreting (References~\cite{biswas:2011} and~\cite{biswas:sandip:2019}). These two extremely inclinedness are not smoothly changed at all. But no change in the physical constrain leads to such a drastically diversified solutions. So, there must exist some ``missing links'' between the two kinds of terminal cases (i.e. adiabatic and MCG flow). If even we are succeeded to find them, what should be the related nature of the density variations and the corresponding thermodynamics is more interesting point. We will try to find out the answers in the subsequent sections.

The rest of the paper is organized as follows. In section~\ref{sec2:profile:fluid}, first we reorganize the structure of the PNF for a rotating BH embedded in quintessence universe. Then we construct the mathematical problem for our model. In subsection~\ref{subsec:speed}, we find solutions for radially inward speed , speed of sound and specific angular momentum as function of radial distance from the BH. We thoroughly analyze these curves as well. In subsection~\ref{subsec:density}, we find the variation of densities of accretion and wind for different parameters and have explained them. Subsection~\ref{subsec:viscosity} deals with the study of the ratio of shear viscosity to entropy density for our model. In section~\ref{sec:conclusion}, we conclude in brief.

\section{Mathematical Modeling of Viscous Accretion onto Rotating SMBHs}
\label{sec2:profile:fluid}

\subsection{Equations for Pseudo Newtonian Force}
We keep effects of different sectors like quintessence etc and construct the PNF~\cite{sarkar:biswas:2019}. We will fix units of length and speed to be $G_NM/c^2$ and $c$ respectively, where $M$ is the mass of the central object and $c$ is the speed of light. We prepare dimensionless parameters $x= \frac{r}{G_NM/c^2}$ and $a = j/c$.

Assuming
\[
\zeta(x) = a\mathcal{A}_q x^{3\omega_q} ~~~~ and
\]

\[
\eta(x)=x^{3(\omega_q-1)} ~~~~ along~ with
\]

$$ 
\alpha(x) = \zeta(x) \{3\omega_q(a^2+x^2)+ 3x^2-8x+a^2\}~,
$$
$$
\beta(x) = (2a^2+6x-8) \zeta^2(x)(A^{-2}_q/a) - 2aA^2_q x ~~ and
$$

$$
\gamma(x)= 2\eta(x)\{(x^2-2x+a^2)x^{3\omega_q} -\mathcal{A}_q x\}^2  \{\mathcal{A}_q (3\omega_q+1) +2\zeta(x)/(a\mathcal{A}_q)\}~~~~,$$
we construct the numerator of the PNF as found by~\cite{sarkar:biswas:2019}
$$
N(x) = \{\alpha(x)+\beta(x) - \sqrt{\gamma(x)}\}^2~.
$$
Next we again assume 

$$
\phi(x) = a \zeta(x)(3\omega_q+1) + 2a^2 x^{6\omega_q}~~ and
$$

$$
\psi(x) = \frac{\zeta(x)}{a}\left[\frac{1}{x^{3}\eta(x)}+ \frac{(2-x)}{\mathcal{A}_q}\right]
$$

and the denominator of the PNF is formed as~\cite{sarkar:biswas:2019}
$$
D(x) = x^3 \{\phi(x) + 2x \psi^2(x)\}^2~.
$$
Finally, we write our PNF as 
\begin{equation}
F_g(x) = N(x)/D(x)
\label{eqn:pnf}
\end{equation}

\subsection{Sound and Fluid Speed Equations}

In this subsection, we will construct the mathematical model from references~\cite{BM:2003} and~\cite{biswas:2011:PNF}. First we will consider the continuity equation,

\[
\frac{\partial \rho}{\partial z} + \vec{\nabla}. (\rho \vec{V}) = 0~~~~,
\]
which is simplified to
\begin{equation}
\frac{d}{dx} \left(x u \Sigma\right) = 0
\label{eqn:math1}
\end{equation}

for stationary and cylindrical structure, where $\Sigma$ is vertically integrated density expressed as 

\begin{equation}
\Sigma = I_C \rho_e h(x),
\label{eqn:math2}
\end{equation}

with $I_C$ = constant (related to EoS of accreting fluid) = 1 (for simplicity), $\rho_e$ = density of the accreting fluid at the equatorial plane, $h(x)$ = half thickness of the disc. $u= u_x = \frac{v_x}{c}$, $v_x$ is the radially inward speed of accretion. Next, we will consider the radial component of stationary Navier Stokes equation.

\begin{equation}
\rho(\vec{V}.\vec{\nabla})\vec{V} = \vec{\nabla}\rho + \rho \gamma \nabla^2 \vec{u} - \vec{F}_{GX}
\end{equation} 

as,

\begin{equation}
u\frac{du}{dx} + \frac{1}{\rho} \frac{dp}{dx}-\frac{\lambda^2}{x^3}+F_g\left( x\right)=0~,
\label{eqn:math3}
\end{equation}
where $F_g(x)$ is the radially inward component of the gravitational force. We use equation (\ref{eqn:pnf}) for $F_g(x)$. 

The azimuthal momentum balance equation turns to be
\begin{equation}
u\frac{d\lambda}{dx}=\frac{1}{x\sum}\frac{d}{dx}\left[x^2\alpha_{SS}\left(P+\rho u^2\right)h(x)\right]~~~~.
\end{equation}
Assuming the vertical equilibrium from the vertical component, we get,

\begin{equation}
h(x) = c_s \sqrt{\frac{x}{F_g}}.
\label{eqn:math4}
\end{equation}

Assume $\Psi =c^2_s + (\alpha-c^2_s)n + \alpha $ and $\mu(x) = (3 - \frac{x}{F_g} \cdot \frac{dF_g}{dx}) $ and the radial inward speed, sonic speed and angular momentum gradients turn to be,

\begin{equation}\label{Ritz_raidal}
\frac{du}{dx} = \frac{u\left[\{\lambda^2-x^3 F_g(x)\}\Psi + x^3\mu(x)c^4_s\right]}{x^3[\Psi u^2 - 2 c^4_s]}~~~~,
\end{equation}

\begin{equation}
\frac{dc_s}{dx} = \left(\frac{1}{2} \mu(x) + \frac{1}{u}\frac{du}{dx}\right) \left\{\frac{(n+1)c_s (c^2-\alpha)}{\Psi}\right\}~~~~and
\end{equation}

\begin{dmath}
\label{differential equation for lambda}
\frac{d \lambda}{d{x}} = \frac{{x} \alpha_{ss}}{u} \left[ \frac{1}{2} \left( \frac{5}{{x}} - \frac{1}{F_g} \frac{dF_g}{d{x}} \right) \left\lbrace \frac{\left(n+1 \right) \alpha - c_s^2}{n} + u^2 \right\rbrace \right.
\left. + 2 u \frac{du}{d{x}} + \left\lbrace \left( \frac{\left(n+1 \right) \alpha - c_s^2}{n} + u^2 \right)  \frac{1}{c_s} -\left( c_s^2 +u^2 \right) \left( \frac{1}{n+1} \frac{2 c_s}{c_s^2 - \alpha} \right) \right\rbrace \frac{dc_s}{d{x}} \right]~~~~. 
\end{dmath}
From the structure of the denominator of the equation (\ref{Ritz_raidal}), it is clear that this will vanish in the domain $(0,~1)$. Now, as the speed of the accreting fluid is very low when it is far from the gravitating object and is very high, even almost equal to the speed of light, i.e., equals to $1$ at the vicinity of the event horizon, we can assume that there will exist a radial distance $x=x_c$, where the vanishing of the denominator takes place. We call this point a critical point. As the flow should be physical, at $x_c$ the numerator should vanish as well and use of L'Hospital's rule will provide us a quadratic equation of the radial velocity gradient and this will generate two different branch of flow : one is accretion and another is expressing the wind. Below, we write the quadratic as
\begin{equation}\label{accretioncqg11}
\mathcal{A}\left(\frac{du}{dx}\right)^{2}_{x=x_c}+\mathcal{B}\left(\frac{du}{dx}\right)_{x=x_c}+\mathcal{C}=0,
\end{equation}
where $$\mathcal{A}=2\left[1-\frac{2\left(c_{sc}^{2}-\alpha\right)\left(n+1\right)\left\{\left(1-n\right)c_{sc}^{2}+2\alpha\left(n+1\right)\right\}}{\left\{\left(1-n\right)c_{sc}^{2}+\alpha\left(n+1\right)\right\}^{2}}\right],$$
\begin{equation}\label{accretioncqg12}
\mathcal{B}=-\frac{2}{c_{sc}^{4}}\frac{\left(c_{sc}^{2}-\alpha\right)\left(n+1\right)\left\{\left(1-n\right)c_{sc}^{2}+2\alpha\left(n+1\right)\right\}}{\left\{\left(1-n\right)c_{sc}^{2}+\alpha\left(n+1\right)\right\}}\left[F_{g}(x_{c})-\frac{\lambda^{2}}{x_{c}^{3}}\right],
\end{equation}

\begin{eqnarray}
\mathcal{C} & = & \left\{\frac{3\lambda^{2}}{x_{c}^{4}}-\left(\frac{dF_{g}}{dx}\right)_{x=x_{c}}\right\}-\left[\left\{\frac{1}{F_{g}}\left(\frac{dF_{g}}{dx}\right)^{2}\right\}_{x=x_{c}}-\frac{3}{x_{c}^{2}}-\left(\frac{1}{F_{g}}\frac{d^{2}F_{g}}{dx^{2}}\right)_{x=x_{c}}\right]\frac{u_{c}^{2}}{2} \\ \nonumber
& & -\frac{u_{c}^{2}}{2c_{sc}^{8}}\left[\left(c_{sc}^{2}-\alpha\right)\left(n+1\right)\left\{\left(1-n\right)c_{sc}^{2}+2\alpha\left(n+1\right)\right\}\right]\left[F_{g}(x_{c})-\frac{\lambda^{2}}{x_{c}^{3}}\right]^{2},
\end{eqnarray}


where $u_c$ is the value of radial velocity at $x=x_c$ and $c_{sc}$ is the speed of sound at $x=x_c$.

\section{Solutions and Analysis}

In this section, we will study different accretion properties, viz, fluid speed, sonic speed, $\lambda/\lambda_k$ ratio, accretion/wind fluid density and $\eta/s$ ratio. We have divided the whole study into three subsections: (i) speeds, (ii) density and (iii) $\eta/s$ ratio.

\subsection{Profiles for Accreting Fluid Speed}
\label{subsec:speed}

\begin{figure}
	\centering
	\setcounter{subfigure}{0}
	\subfigure[$\Gamma=1.6,    \alpha_{ss}=10^{-4}$]{\includegraphics[width=0.24\textwidth]{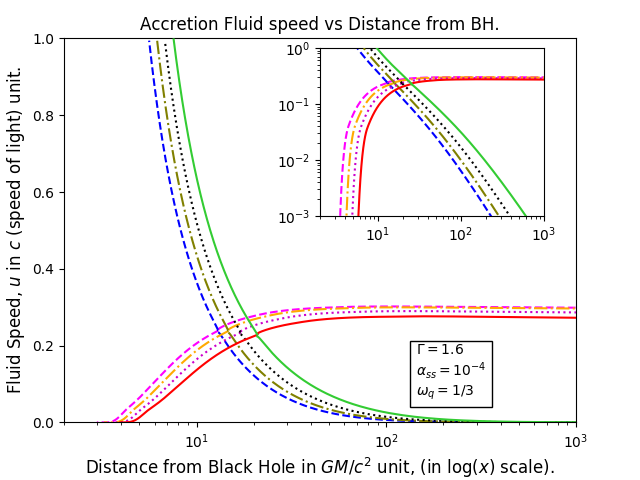}} 
	\subfigure[$\Gamma=1.6,    \alpha_{ss}=10^{-2}$]{\includegraphics[width=0.24\textwidth]{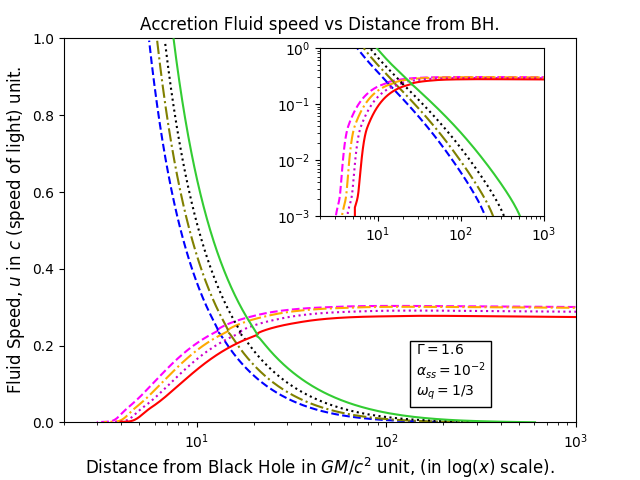}} 
	\subfigure[$\Gamma=0.09,    \alpha_{ss}=10^{-4}$]{\includegraphics[width=0.24\textwidth]{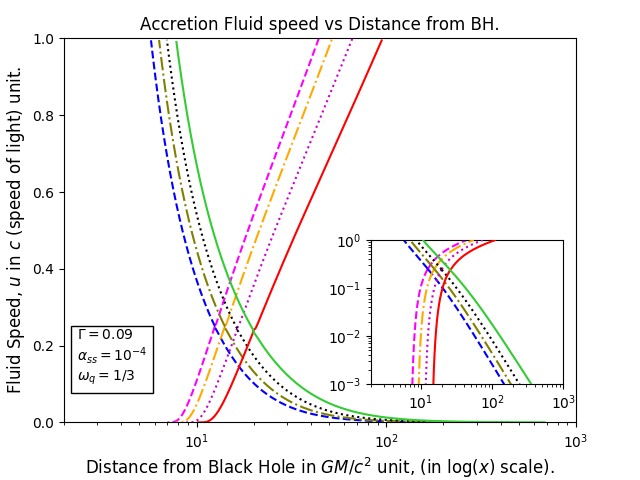}}
	\subfigure[$\Gamma=0.09,    \alpha_{ss}=10^{-2}$]{\includegraphics[width=0.24\textwidth]{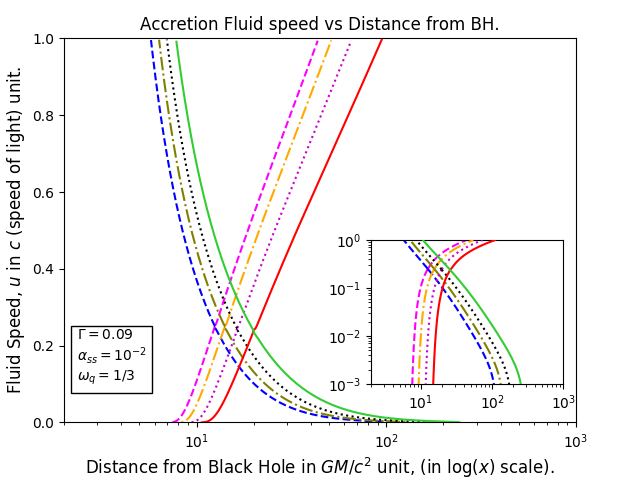}}
	
	\caption*{\textbf{\emph{Figure 1.1:}} Images for $\lambda_c=2.7, \omega_q = 1/3, A_q=0.01$. Red solid line shows wind for $a=0$,  Green solid line shows accretion for $a=0$, Purple dotted line shows wind for $a=0.5$,  Black dotted line shows accretion for $a=0.5$, Orange dash-dotted line shows wind for $a=0.9$,  Olive dash-dotted line shows accretion for $a=0.9$ and Pink dashed-dashed line shows wind for $a=0.998$,  Blue dashed-dashed line shows accretion for $a=0.998$}
	\setcounter{subfigure}{0}
	\subfigure[$\Gamma=1.6,    \alpha_{ss}=10^{-4}$]{\includegraphics[width=0.24\textwidth]{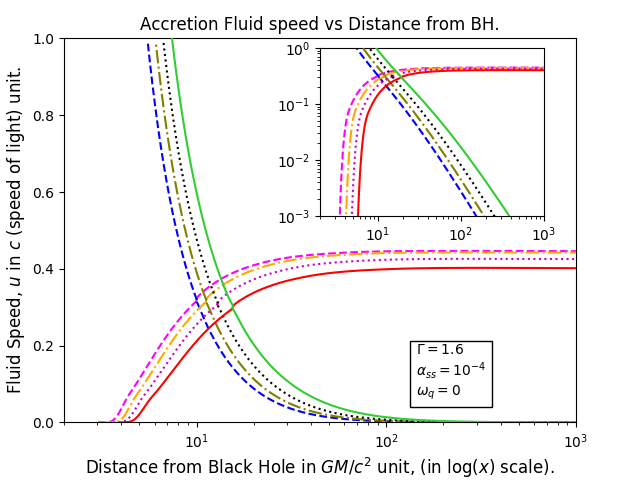}} 
	\subfigure[$\Gamma=1.6,    \alpha_{ss}=10^{-2}$]{\includegraphics[width=0.24\textwidth]{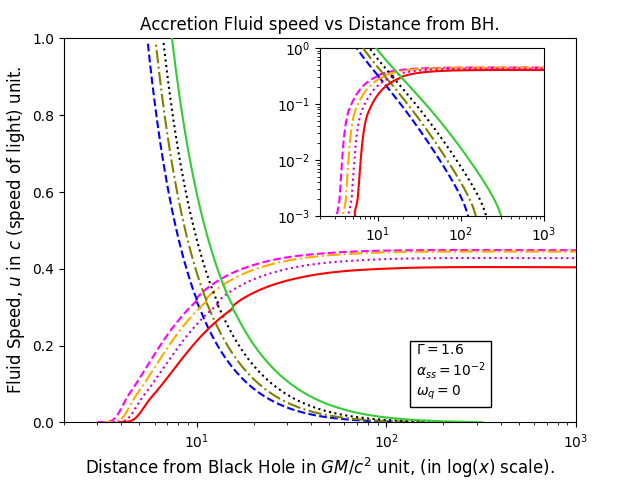}} 
	\subfigure[$\Gamma=0.09,    \alpha_{ss}=10^{-4}$]{\includegraphics[width=0.24\textwidth]{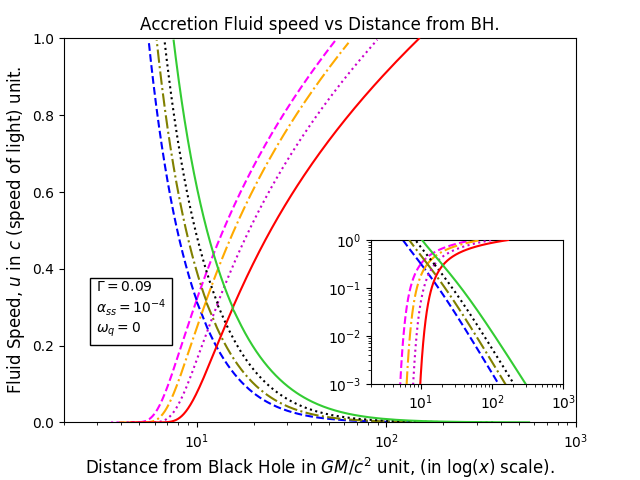}}
	\subfigure[$\Gamma=0.09,    \alpha_{ss}=10^{-2}$]{\includegraphics[width=0.24\textwidth]{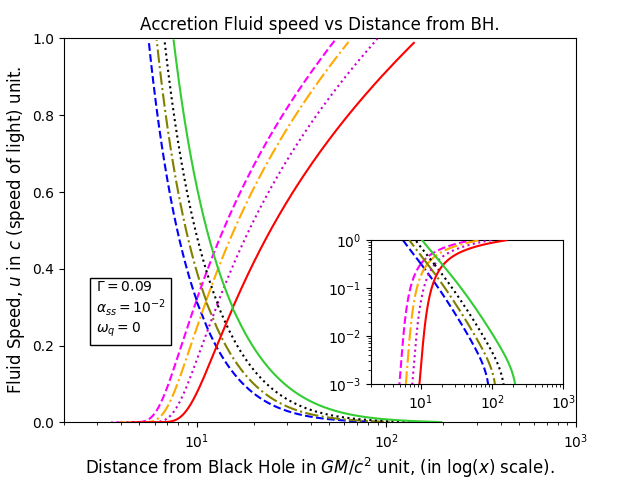}}
	\caption*{\textbf{\emph{Figure 1.2:}} Images for $\lambda_c=2.7, \omega_q = 0, A_q=0.01$. Red solid line shows wind for $a=0$,  Green solid line shows accretion for $a=0$, Purple dotted line shows wind for $a=0.5$,  Black dotted line shows accretion for $a=0.5$, Orange dash-dotted line shows wind for $a=0.9$,  Olive dash-dotted line shows accretion for $a=0.9$ and Pink dashed-dashed line shows wind for $a=0.998$,  Blue dashed-dashed line shows accretion for $a=0.998$}
	\setcounter{subfigure}{0}
	\subfigure[$\Gamma=1.6,    \alpha_{ss}=10^{-4}$]{\includegraphics[width=0.24\textwidth]{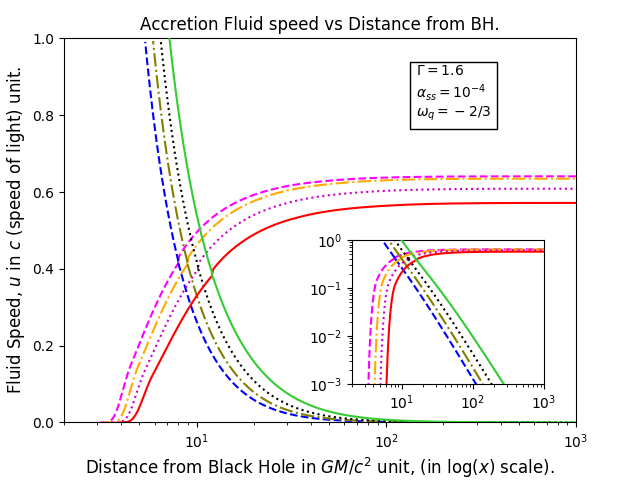}} 
	\subfigure[$\Gamma=1.6,    \alpha_{ss}=10^{-2}$]{\includegraphics[width=0.24\textwidth]{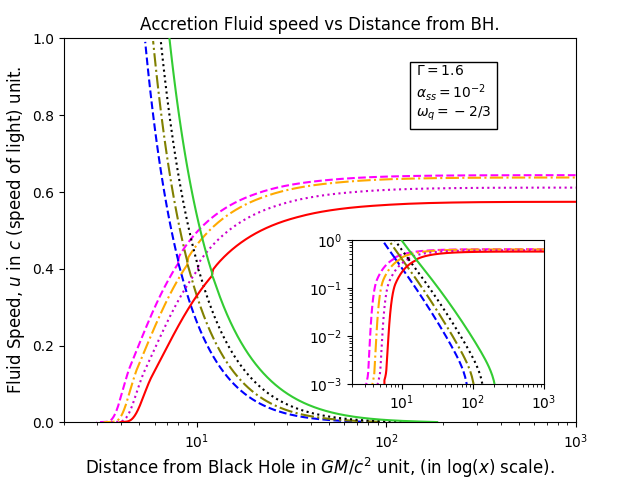}} 
	\subfigure[$\Gamma=0.09,    \alpha_{ss}=10^{-4}$]{\includegraphics[width=0.24\textwidth]{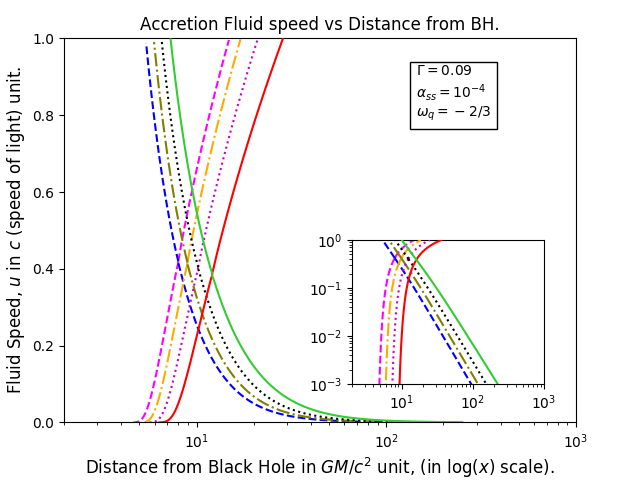}}
	\subfigure[$\Gamma=0.09,    \alpha_{ss}=10^{-2}$]{\includegraphics[width=0.24\textwidth]{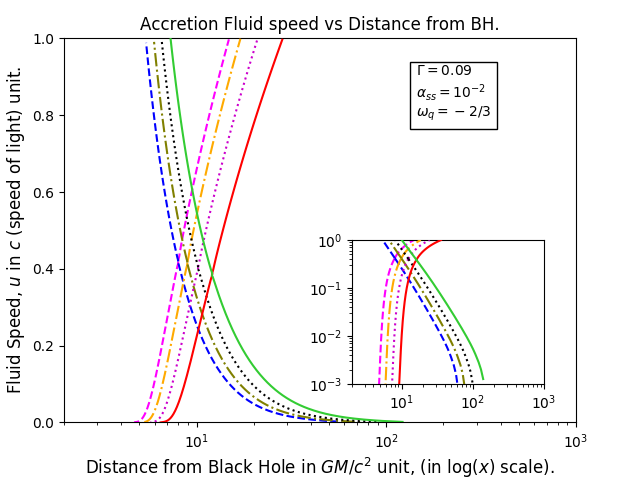}}
	\caption*{\textbf{\emph{Figure 1.3:}} Images for $\lambda_c=2.7, \omega_q = -2/3, A_q=10^{-10}$. Red solid line shows wind for $a=0$,  Green solid line shows accretion for $a=0$, Purple dotted line shows wind for $a=0.5$,  Black dotted line shows accretion for $a=0.5$, Orange dash-dotted line shows wind for $a=0.9$,  Olive dash-dotted line shows accretion for $a=0.9$ and Pink dashed-dashed line shows wind for $a=0.998$,  Blue dashed-dashed line shows accretion for $a=0.998$}
	\caption{Curves for fluid speed vs radial distance from the BH.}
\end{figure}

\begin{figure}
	\ContinuedFloat
	\setcounter{subfigure}{0}
	\subfigure[$\Gamma=1.6,    \alpha_{ss}=10^{-4}$]{\includegraphics[width=0.24\textwidth]{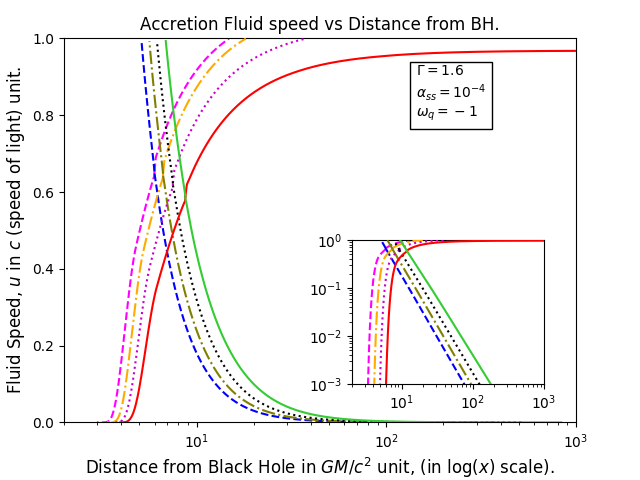}} 
	\subfigure[$\Gamma=1.6,    \alpha_{ss}=10^{-2}$]{\includegraphics[width=0.24\textwidth]{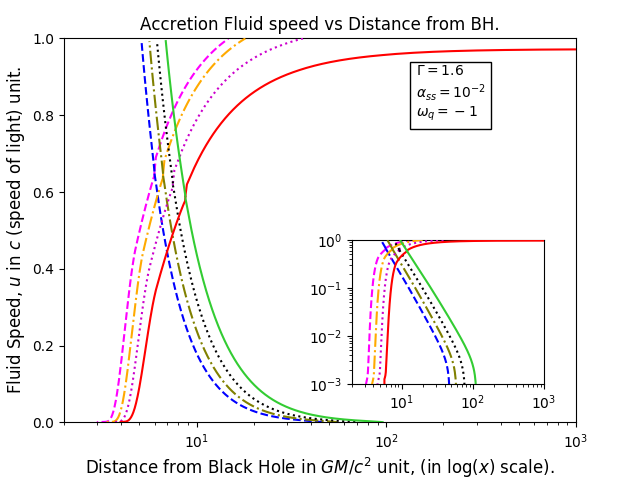}} 
	\subfigure[$\Gamma=0.09,    \alpha_{ss}=10^{-4}$]{\includegraphics[width=0.24\textwidth]{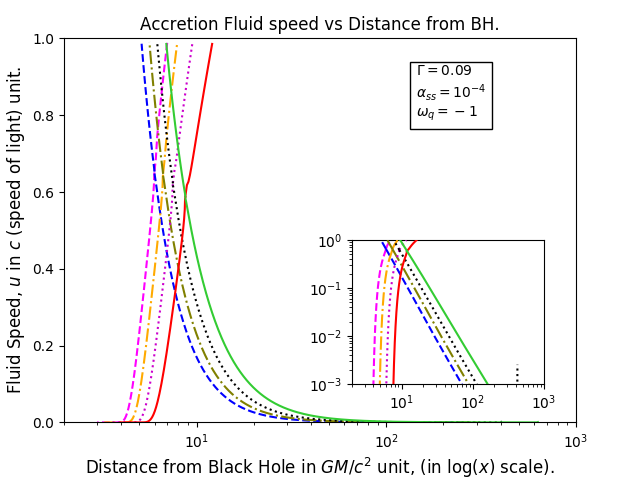}}
	\subfigure[$\Gamma=0.09,    \alpha_{ss}=10^{-2}$]{\includegraphics[width=0.24\textwidth]{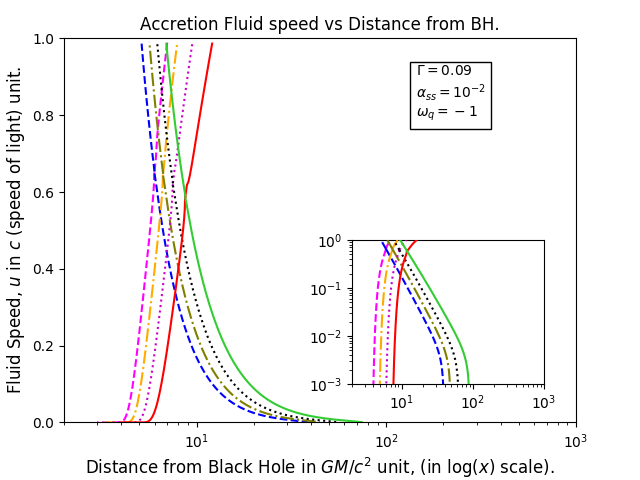}}
	\caption*{\textbf{\emph{Figure 1.4:}} Images for $\lambda_c=2.7, \omega_q = -1, A_q=10^{-10}$. Red solid line shows wind for $a=0$,  Green solid line shows accretion for $a=0$, Purple dotted line shows wind for $a=0.5$,  Black dotted line shows accretion for $a=0.5$, Orange dash-dotted line shows wind for $a=0.9$,  Olive dash-dotted line shows accretion for $a=0.9$ and Pink dashed-dashed line shows wind for $a=0.998$,  Blue dashed-dashed line shows accretion for $a=0.998$}
	\caption{Curves for fluid speed vs radial distance from the BH.}
	\label{fig:fluid}
\end{figure}

\begin{figure}
	\centering
	\setcounter{subfigure}{0}
	\subfigure[$\Gamma=1.6,    \alpha_{ss}=10^{-4}$]{\includegraphics[width=0.24\textwidth]{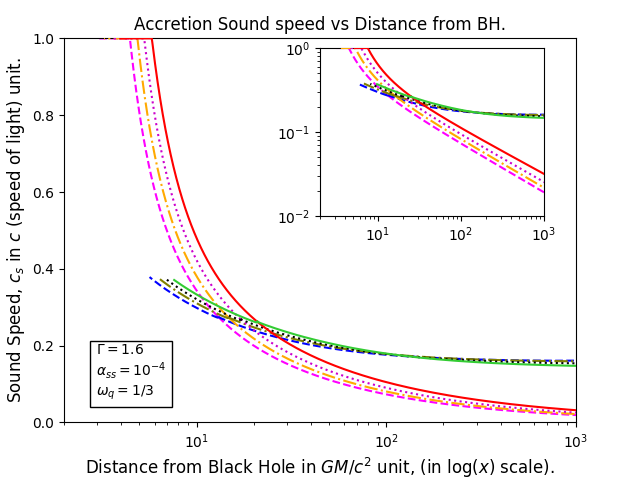}} 
	\subfigure[$\Gamma=1.6,    \alpha_{ss}=10^{-2}$]{\includegraphics[width=0.24\textwidth]{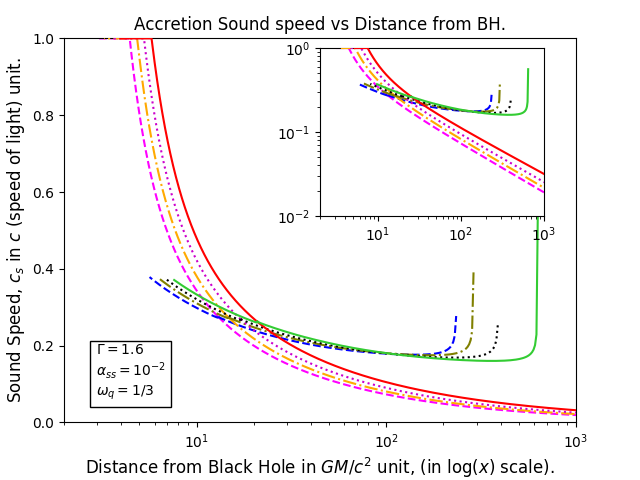}} 
	\subfigure[$\Gamma=0.09,    \alpha_{ss}=10^{-4}$]{\includegraphics[width=0.24\textwidth]{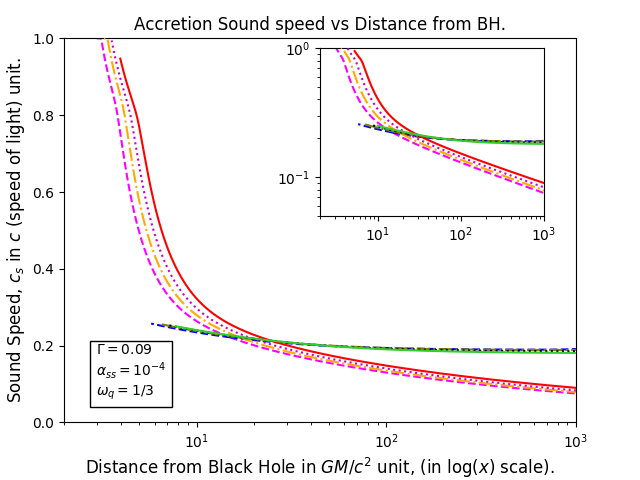}}
	\subfigure[$\Gamma=0.09,    \alpha_{ss}=10^{-2}$]{\includegraphics[width=0.24\textwidth]{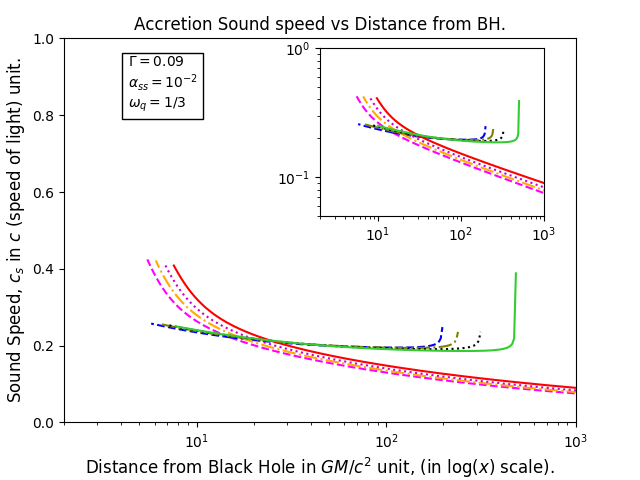}}
	\caption*{\textbf{\emph{Figure 2.1:}} Images for $\lambda_c=2.7, \omega_q = 1/3, A_q=0.01$. Red solid line shows wind for $a=0$,  Green solid line shows accretion for $a=0$, Purple dotted line shows wind for $a=0.5$,  Black dotted line shows accretion for $a=0.5$, Orange dash-dotted line shows wind for $a=0.9$,  Olive dash-dotted line shows accretion for $a=0.9$ and Pink dashed-dashed line shows wind for $a=0.998$,  Blue dashed-dashed line shows accretion for $a=0.998$}
	\setcounter{subfigure}{0}
	\subfigure[$\Gamma=1.6,    \alpha_{ss}=10^{-4}$]{\includegraphics[width=0.24\textwidth]{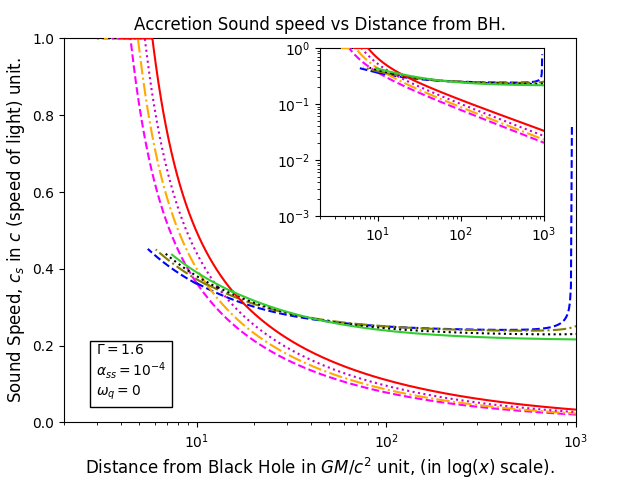}} 
	\subfigure[$\Gamma=1.6,    \alpha_{ss}=10^{-2}$]{\includegraphics[width=0.24\textwidth]{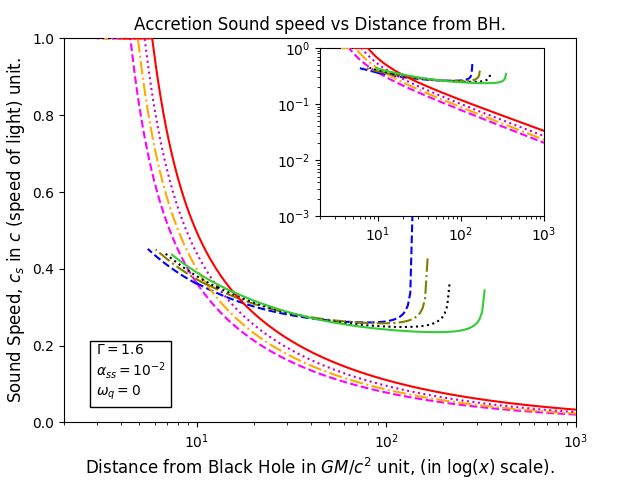}} 
	\subfigure[$\Gamma=0.09,    \alpha_{ss}=10^{-4}$]{\includegraphics[width=0.24\textwidth]{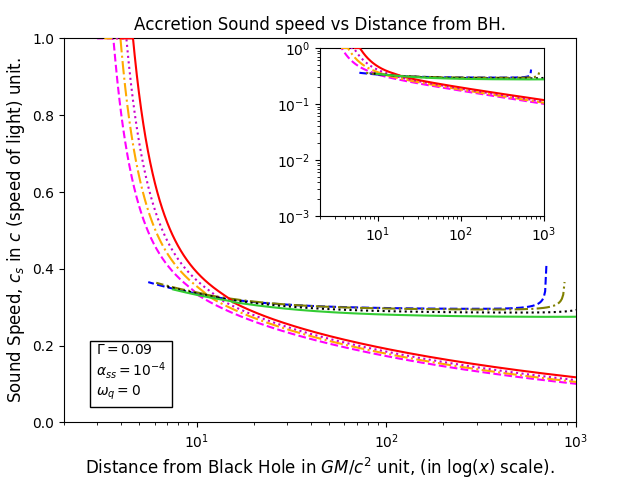}}
	\subfigure[$\Gamma=0.09,    \alpha_{ss}=10^{-2}$]{\includegraphics[width=0.24\textwidth]{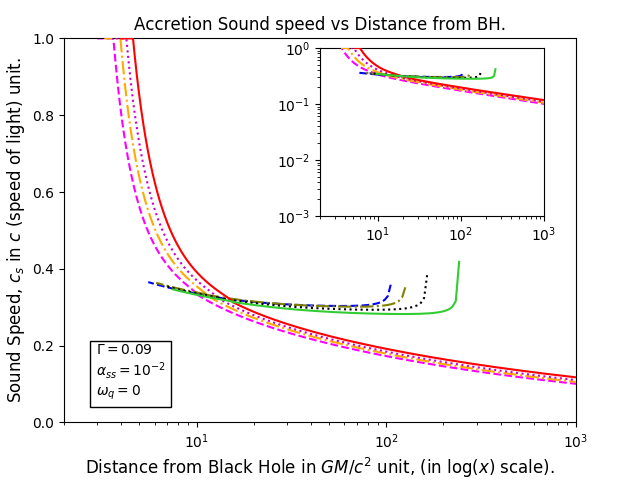}}
	\caption*{\textbf{\emph{Figure 2.2:}} Images for $\lambda_c=2.7, \omega_q = 0, A_q=0.01$. Red solid line shows wind for $a=0$,  Green solid line shows accretion for $a=0$, Purple dotted line shows wind for $a=0.5$,  Black dotted line shows accretion for $a=0.5$, Orange dash-dotted line shows wind for $a=0.9$,  Olive dash-dotted line shows accretion for $a=0.9$ and Pink dashed-dashed line shows wind for $a=0.998$,  Blue dashed-dashed line shows accretion for $a=0.998$}
	\setcounter{subfigure}{0}
	\subfigure[$\Gamma=1.6,    \alpha_{ss}=10^{-4}$]{\includegraphics[width=0.24\textwidth]{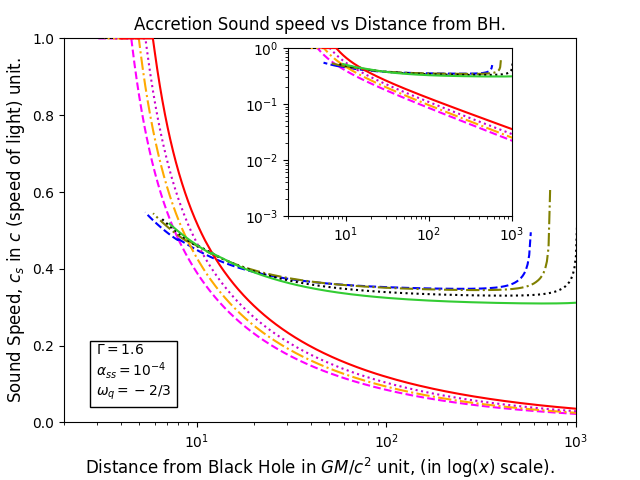}} 
	\subfigure[$\Gamma=1.6,    \alpha_{ss}=10^{-2}$]{\includegraphics[width=0.24\textwidth]{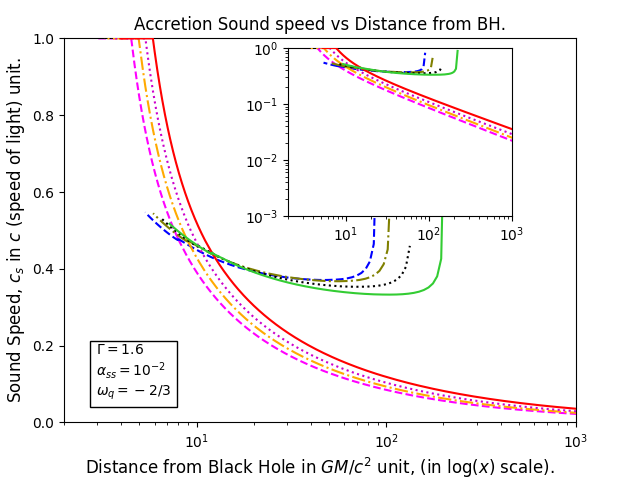}} 
	\subfigure[$\Gamma=0.09,    \alpha_{ss}=10^{-4}$]{\includegraphics[width=0.24\textwidth]{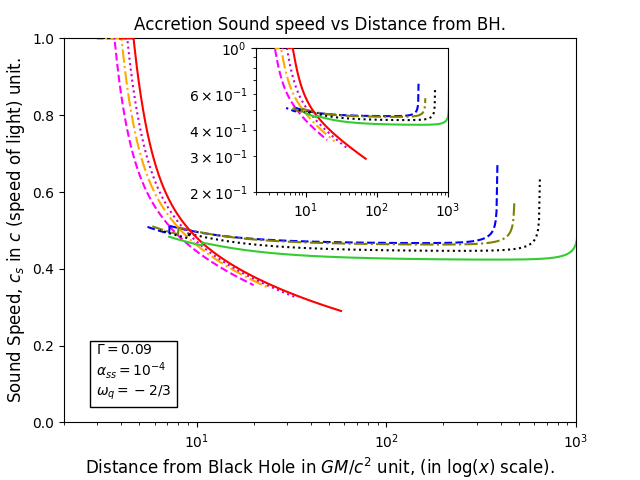}}
	\subfigure[$\Gamma=0.09,    \alpha_{ss}=10^{-2}$]{\includegraphics[width=0.24\textwidth]{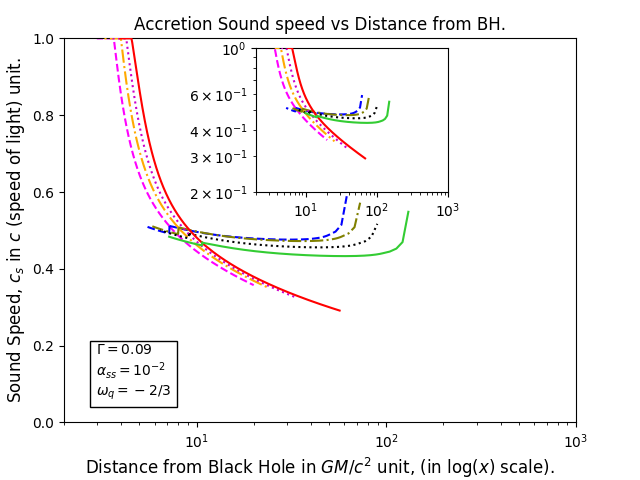}}
	\caption*{\textbf{\emph{Figure 2.3:}} Images for $\lambda_c=2.7, \omega_q = -2/3, A_q=10^{-10}$. Red solid line shows wind for $a=0$,  Green solid line shows accretion for $a=0$, Purple dotted line shows wind for $a=0.5$,  Black dotted line shows accretion for $a=0.5$, Orange dash-dotted line shows wind for $a=0.9$,  Olive dash-dotted line shows accretion for $a=0.9$ and Pink dashed-dashed line shows wind for $a=0.998$,  Blue dashed-dashed line shows accretion for $a=0.998$}
	\caption{Curves for sound speed vs radial distance from the BH.}
\end{figure}

\begin{figure}
	\ContinuedFloat
	\setcounter{subfigure}{0}
	\subfigure[$\Gamma=1.6,    \alpha_{ss}=10^{-4}$]{\includegraphics[width=0.24\textwidth]{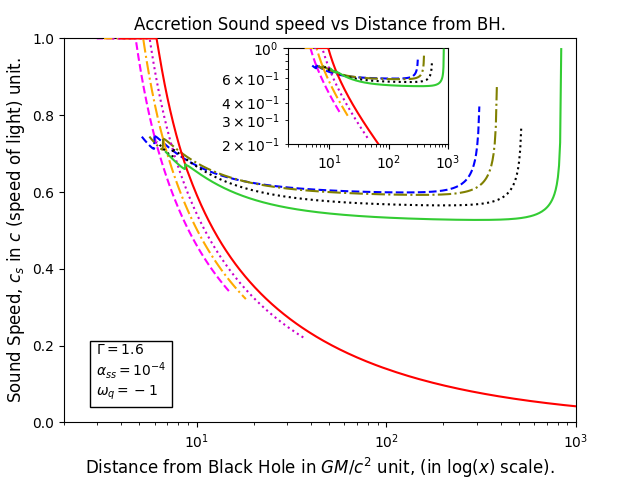}} 
	\subfigure[$\Gamma=1.6,    \alpha_{ss}=10^{-2}$]{\includegraphics[width=0.24\textwidth]{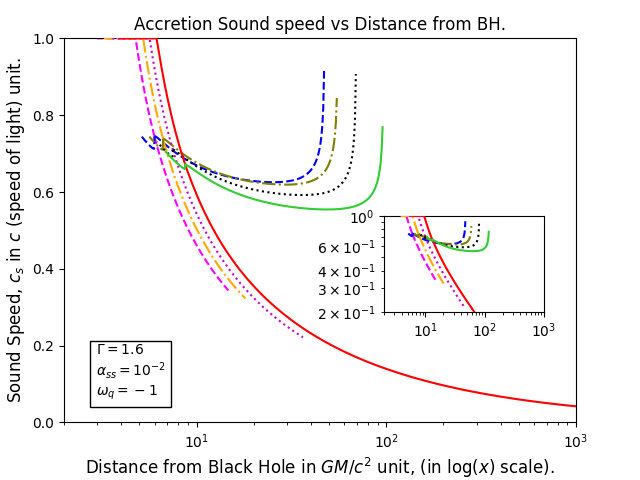}} 
	\subfigure[$\Gamma=0.09,    \alpha_{ss}=10^{-4}$]{\includegraphics[width=0.24\textwidth]{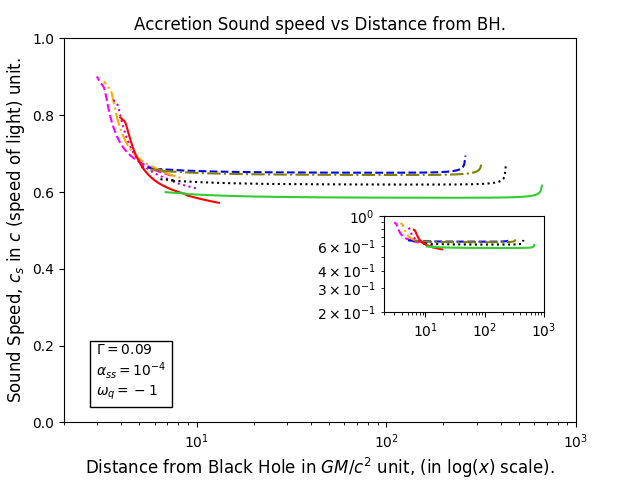}}
	\subfigure[$\Gamma=0.09,    \alpha_{ss}=10^{-2}$]{\includegraphics[width=0.24\textwidth]{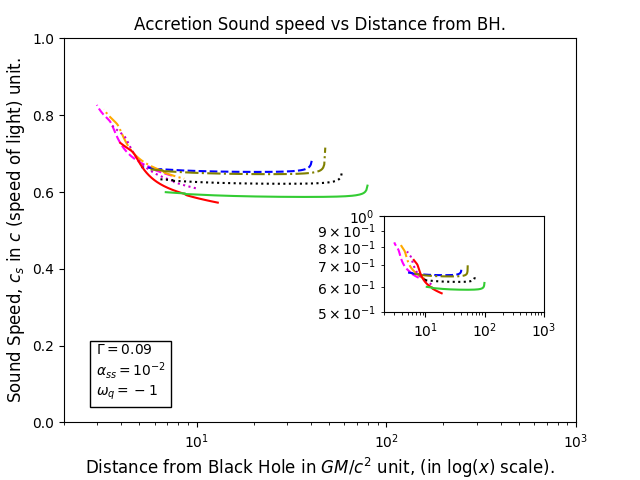}}
	\caption*{\textbf{\emph{Figure 2.4:}} Images for $\lambda_c=2.7, \omega_q = -1, A_q=10^{-10}$. Red solid line shows wind for $a=0$,  Green solid line shows accretion for $a=0$, Purple dotted line shows wind for $a=0.5$,  Black dotted line shows accretion for $a=0.5$, Orange dash-dotted line shows wind for $a=0.9$,  Olive dash-dotted line shows accretion for $a=0.9$ and Pink dashed-dashed line shows wind for $a=0.998$,  Blue dashed-dashed line shows accretion for $a=0.998$}
	\caption{Curves for sound speed vs radial distance from the BH.}
	\label{fig:sound}
\end{figure}

We have plotted figures~\ref{fig:fluid}.1.a to~\ref{fig:fluid}.4.d to show the variations of fluid's radially inward speed w.r.t radial distance. Rows entitled \ref{fig:fluid}.1, \ref{fig:fluid}.2, \ref{fig:fluid}.3 and \ref{fig:fluid}.4 are for $\omega_q=1/3, 0, -2/3$ and $-1$ respectively. Columns entitled (a) and (b) are for adiabatic accretions with polytropic index = 1.6 with viscosity given by $\alpha_{ss}=10^{-4}$ and $10^{-2}$ respectively. Columns entitled (c) and (d) are for adiabatic index 0.09 with viscosity parameter $\alpha_{ss}=10^{-4}$ and $10^{-2}$ respectively. In each figure, solid lines are (green) and wind(red) for $a=0$, dotted lines are accretion(black) and wind(purple) for $a=0.5$, dot-dashed lines are accretion(olive) and wind (orange) for $a=0.9$ and dashed lines are accretion(blue) and wind(pink) for $a=0.998$. Inset contains $\log(u)$ vs. $\log(x)$ curves whereas the offset shows $u$ vs. $log(x)$ variations.

Now we are ready to analyze figure~\ref{fig:fluid}.1.a. The common features of the curves are same. Accretion speed is raising as we move towards the BH. Wind speed is low near the BH. As we go far, it increases and then becomes almost constant. Point to be noted that the accretion becomes fainter as we move far from the BH. This fainting rate increases as the value of the spin parameter increases. The more is the rotation, the nearer the sensible wind flow starts from the BH. As viscosity increases, in figure~\ref{fig:fluid}.1.b, we observe that the accretion to fall abruptly at the distant parts of the accretion disc. So, inclusion of viscosity reduces the angular momentum transport efficiency which ultimately causes the reduction of the physical radius of the disc. This is very much clear in \ref{fig:fluid}.1.d, which is drawn for $\gamma=0.09, \alpha_{ss}=10^{-2}$. We failed to reach to the positive values of $n_{MCG}$ which may indicate the existence of DE accretion onto a quintessence BH. $\Gamma=0.09$ is the lowest value for which get physical solutions. We see wind speed to increase abruptly its value reaches light's speed at finite distance. We conclude that if the value of $\Gamma$ is low, wind dominates accretion. This nature is, however, accompanied by inclusion of high viscosity. For high spin parameter, the radius, where wind speed becomes equal to that of light is less. 

Before approaching to the other cases of $\omega_q$, we try to find out the case for which the accretion stops or the wind reaches the speed of light at a finite distance. To do this, we will look at the sound speed and $\frac{\lambda}{\lambda_k}$ curves which are plotted in figures~\ref{fig:sound}.1.a to~\ref{fig:sound}.4.d and~\ref{fig:lambda}.1.a to \ref{fig:lambda}.4.d respectively.

\begin{figure}
	\centering
	\setcounter{subfigure}{0}
	\subfigure[$\Gamma=1.6,    \alpha_{ss}=10^{-4}$]{\includegraphics[width=0.24\textwidth]{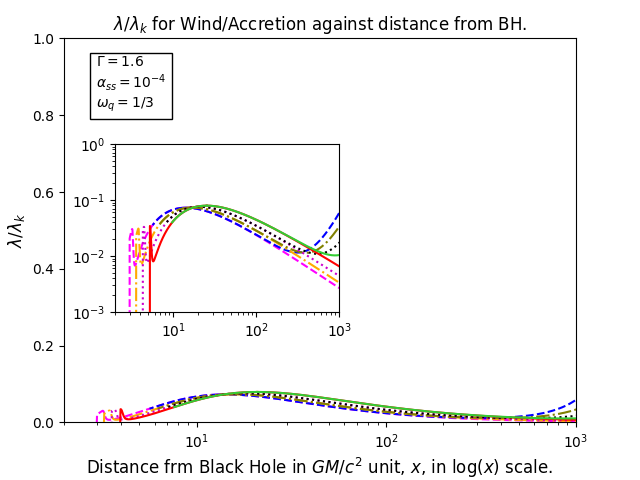}} 
	\subfigure[$\Gamma=1.6,    \alpha_{ss}=10^{-2}$]{\includegraphics[width=0.24\textwidth]{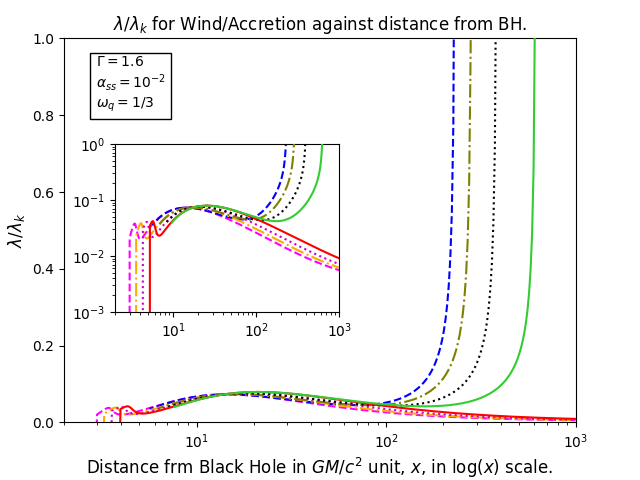}} 
	\subfigure[$\Gamma=0.09,    \alpha_{ss}=10^{-4}$]{\includegraphics[width=0.24\textwidth]{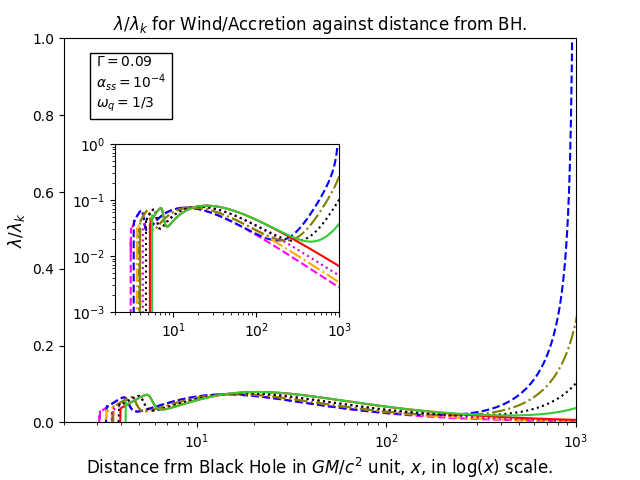}}
	\subfigure[$\Gamma=0.09,    \alpha_{ss}=10^{-2}$]{\includegraphics[width=0.24\textwidth]{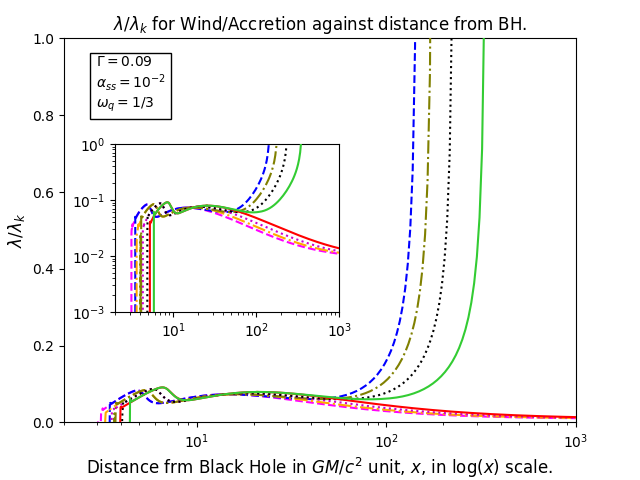}}
	\caption*{\textbf{\emph{Figure 3.1:}} Images for $\lambda_c=2.7, \omega_q = 1/3, A_q=0.01$. Red solid line shows wind for $a=0$,  Green solid line shows accretion for $a=0$, Purple dotted line shows wind for $a=0.5$,  Black dotted line shows accretion for $a=0.5$, Orange dash-dotted line shows wind for $a=0.9$,  Olive dash-dotted line shows accretion for $a=0.9$ and Pink dashed-dashed line shows wind for $a=0.998$,  Blue dashed-dashed line shows accretion for $a=0.998$}
	\setcounter{subfigure}{0}
	\subfigure[$\Gamma=1.6,    \alpha_{ss}=10^{-4}$]{\includegraphics[width=0.24\textwidth]{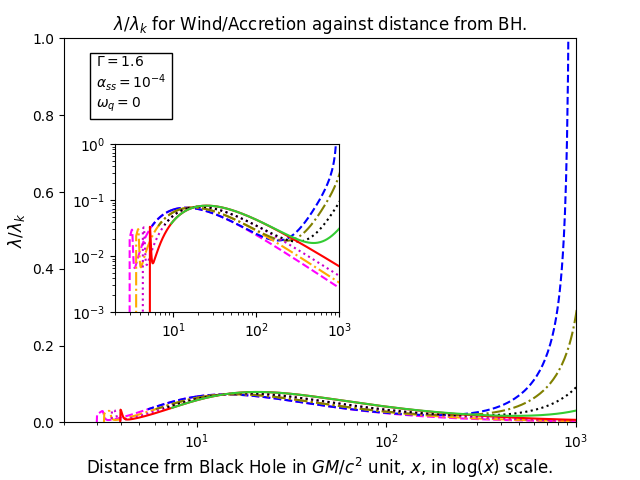}} 
	\subfigure[$\Gamma=1.6,    \alpha_{ss}=10^{-2}$]{\includegraphics[width=0.24\textwidth]{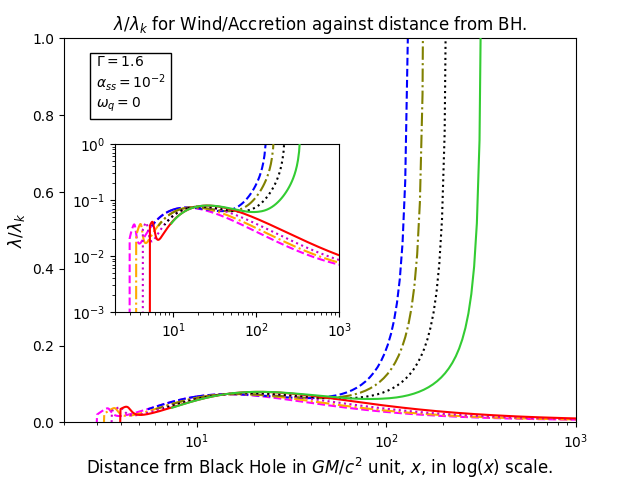}} 
	\subfigure[$\Gamma=0.09,    \alpha_{ss}=10^{-4}$]{\includegraphics[width=0.24\textwidth]{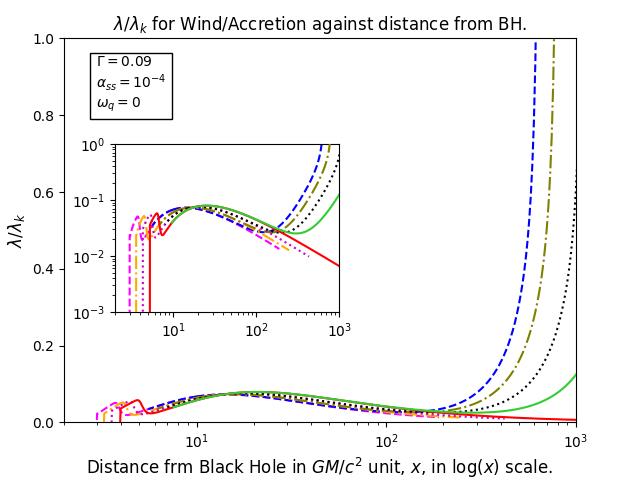}}
	\subfigure[$\Gamma=0.09,    \alpha_{ss}=10^{-2}$]{\includegraphics[width=0.24\textwidth]{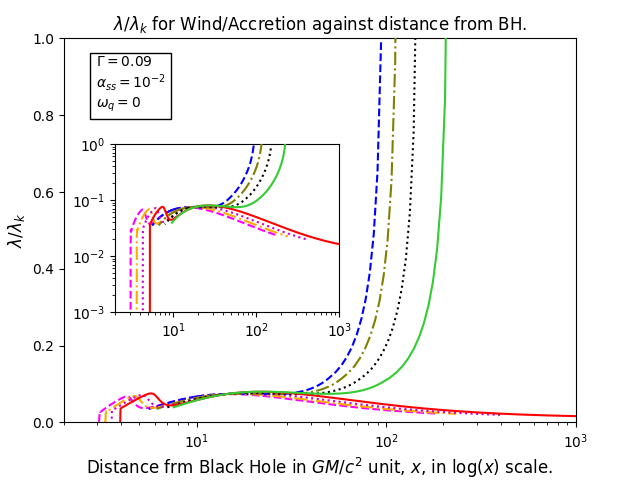}}
	\caption*{\textbf{\emph{Figure 3.2:}} Images for $\lambda_c=2.7, \omega_q = 0, A_q=10.01$. Red solid line shows wind for $a=0$,  Green solid line shows accretion for $a=0$, Purple dotted line shows wind for $a=0.5$,  Black dotted line shows accretion for $a=0.5$, Orange dash-dotted line shows wind for $a=0.9$,  Olive dash-dotted line shows accretion for $a=0.9$ and Pink dashed-dashed line shows wind for $a=0.998$,  Blue dashed-dashed line shows accretion for $a=0.998$}
	\setcounter{subfigure}{0}
	\subfigure[$\Gamma=1.6,    \alpha_{ss}=10^{-4}$]{\includegraphics[width=0.24\textwidth]{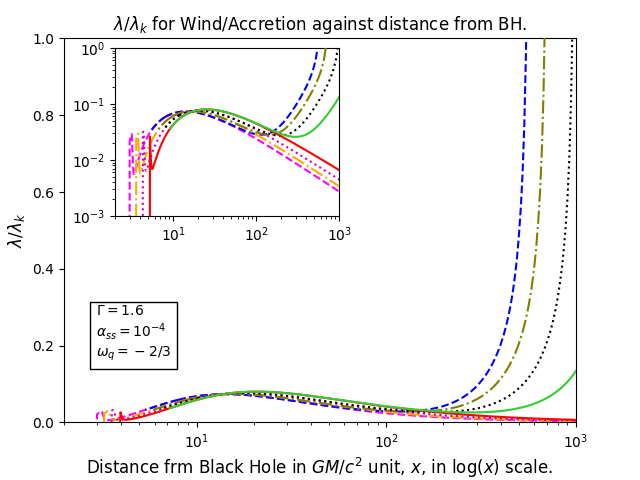}} 
	\subfigure[$\Gamma=1.6,    \alpha_{ss}=10^{-2}$]{\includegraphics[width=0.24\textwidth]{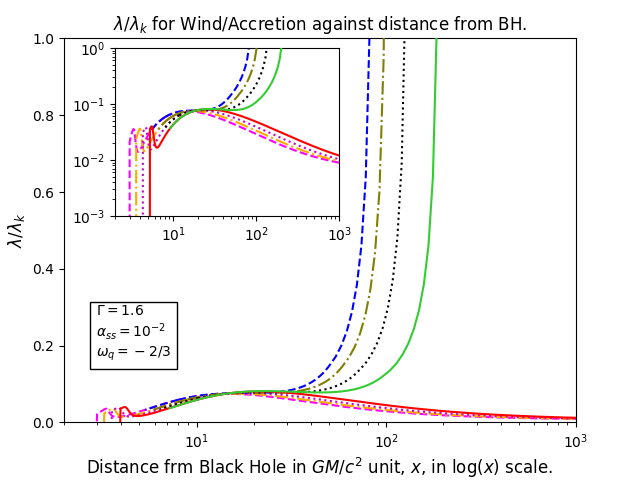}} 
	\subfigure[$\Gamma=0.09,    \alpha_{ss}=10^{-4}$]{\includegraphics[width=0.24\textwidth]{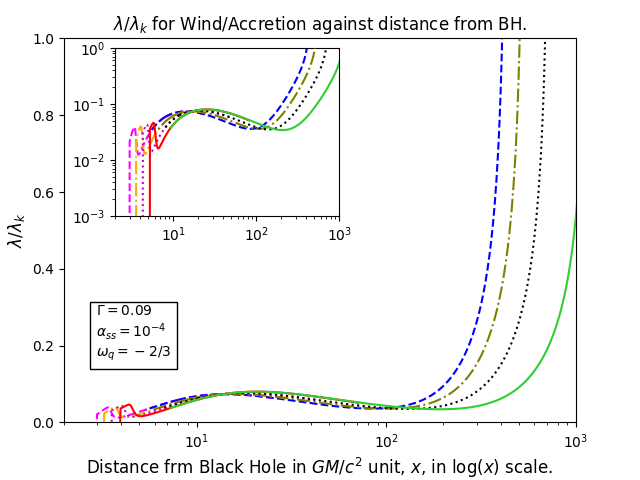}}
	\subfigure[$\Gamma=0.09,    \alpha_{ss}=10^{-2}$]{\includegraphics[width=0.24\textwidth]{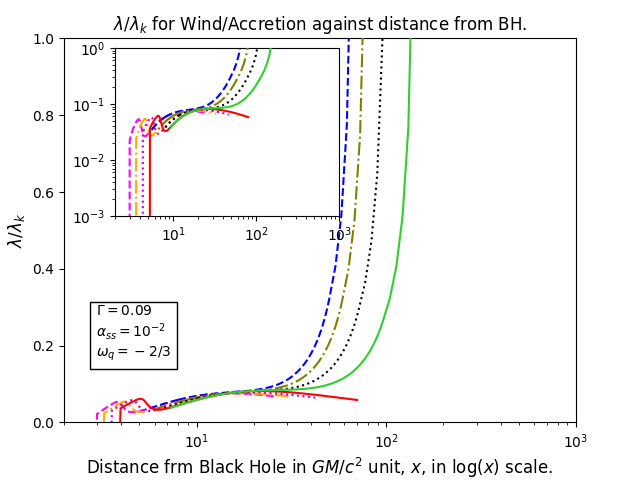}}
	\caption*{\textbf{\emph{Figure 3.3:}} Images for $\lambda_c=2.7, \omega_q = -2/3, A_q=10^{-10}$. Red solid line shows wind for $a=0$,  Green solid line shows accretion for $a=0$, Purple dotted line shows wind for $a=0.5$,  Black dotted line shows accretion for $a=0.5$, Orange dash-dotted line shows wind for $a=0.9$,  Olive dash-dotted line shows accretion for $a=0.9$ and Pink dashed-dashed line shows wind for $a=0.998$,  Blue dashed-dashed line shows accretion for $a=0.998$}
	\caption{$\lambda/\lambda_k$ vs radial distance plotting for accretion and wind branches for different parameters.}
\end{figure}

\begin{figure}
	\ContinuedFloat
	\setcounter{subfigure}{0}
	\subfigure[$\Gamma=1.6,    \alpha_{ss}=10^{-4}$]{\includegraphics[width=0.24\textwidth]{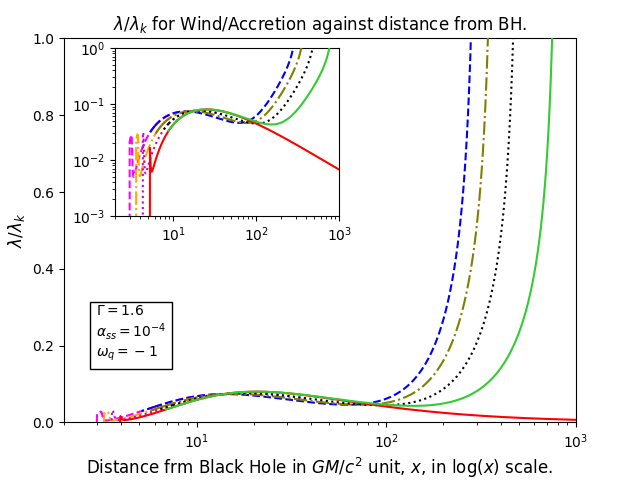}} 
	\subfigure[$\Gamma=1.6,    \alpha_{ss}=10^{-2}$]{\includegraphics[width=0.24\textwidth]{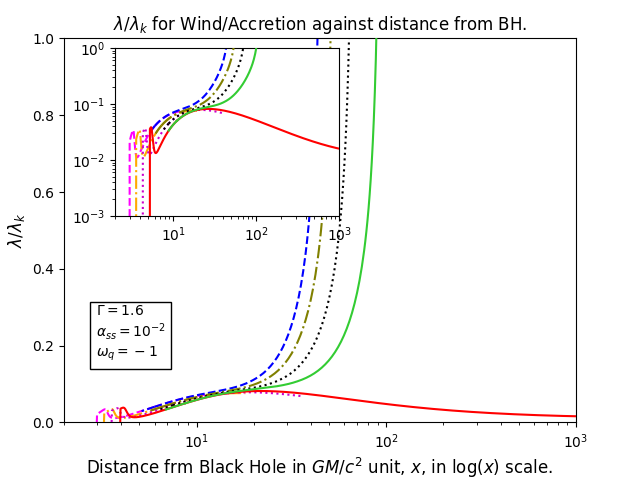}} 
	\subfigure[$\Gamma=0.09,    \alpha_{ss}=10^{-4}$]{\includegraphics[width=0.24\textwidth]{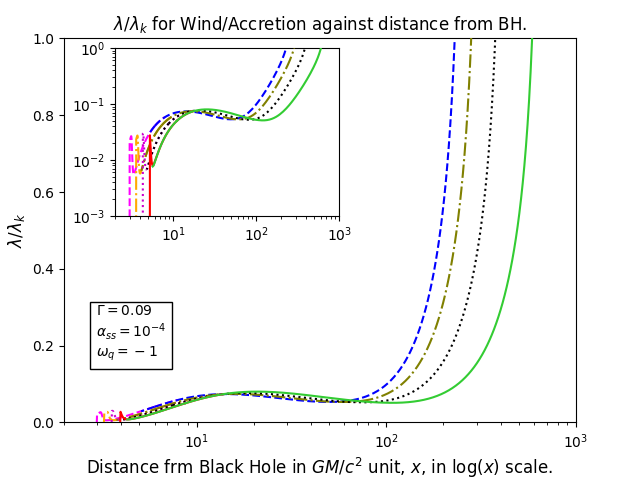}}
	\subfigure[$\Gamma=0.09,    \alpha_{ss}=10^{-2}$]{\includegraphics[width=0.24\textwidth]{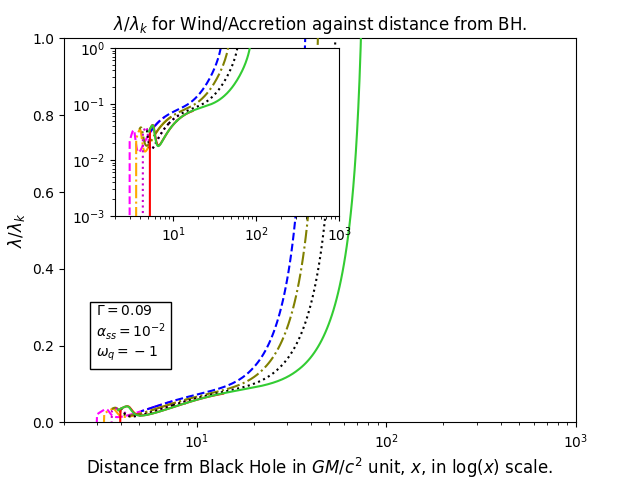}}
	\caption*{\textbf{\emph{Figure 3.4:}} Images for $\lambda_c=2.7, \omega_q = -1, A_q=10^{-10}$. Red solid line shows wind for $a=0$,  Green solid line shows accretion for $a=0$, Purple dotted line shows wind for $a=0.5$,  Black dotted line shows accretion for $a=0.5$, Orange dash-dotted line shows wind for $a=0.9$,  Olive dash-dotted line shows accretion for $a=0.9$ and Pink dashed-dashed line shows wind for $a=0.998$,  Blue dashed-dashed line shows accretion for $a=0.998$}
	\caption{$\lambda/\lambda_k$ plotting for Accretion and Wind for different parameters.}
	\label{fig:lambda}
\end{figure}

For radiation dominated era, we find the sound speed of wind branch decreases slowly as we go far from the BH. As viscosity is considered, the sound speed for the wind branch is found to blow up at a finite distance $x=x_{end}\left(\omega_q, \alpha_{ss}\right)$. Exactly at $x=x_{end}\left(\omega_q\right)$ the fluid speed for wind branch for wind branch is found to be zero. So, at this end  regions of the corresponding disc, the fluid is acting mere stiff which causes the raising in the sound speed's value but lowering in original fluid speed. As $\alpha_{ss}$ increases, the value of $x_{end}$ decreases. Similarly, as $\omega_q$ is reduced, $x_{end}$ decreases as well. So, both the effects of viscosity and quintessential nature of the BH's background shortens the effective disc length. Accretion branch's accretion speed is found to have sound speed which is low at farthest point and decreases as the distance is reduced.

$\lambda/\lambda_k$ curves are saying where the angular momentum is being greater than that possessed by a Keplerian orbit. If $\lambda/\lambda_k > 1$, the disc is rotating with a speed which is even greater than  to  fight the inward gravitational pull. This will  break the structure of stable accretion disc after $x=x_{end}$ and the part beyond it will be truncated off.

These three sets of graphs conclude that a viscous accretion onto a quintessence contaminated BH is weakened/shortened by four factors: the rotation of the BH, negativity of $\omega_q$ in which the BH is embedded, value of $\alpha_{ss}$, i.e., the viscosity imposed and the negativity inserted in EoS of accreting fluid. If all these factors increase, the accretion disc is fainted causing a weaker feeding process.

To concretize these results, in the next subsection, we will study the variation of density of accretion and the wind for all the possible cases.

\subsection{Profiles for Accreting Fluid Density}
\label{subsec:density}

To sustain an accretion process to run, the total system considered must not have luminosity greater than a maximum limit. Beyond this limit, the radiation pressure is so high, that any object will overcome gravitational pull. This will let no matter to fall inward and accretion will stop. This maximum limit is known as Eddington luminosity ($L_{Edd}$, say). To obtain this limit, we balance the gravitational force with radiation force as,

\[
F_{Grav}(M, m, R) = \frac{G_N M m}{R^2} = P_{rad}k m = \frac{L}{c} \cdot \frac{1}{4\pi R^2} k m = \frac{L}{c} \cdot \frac{1}{4\pi R^2} \frac{\sigma_T}{m_p}m
\]

where $k, \sigma_T, m_p,$ and $L$ are opacity, Thompson Scattering cross-section, mass of proton and luminosity respectively. If this $L$ is taken to be equal to Eddington Luminosity, we obtain

\begin{equation}
L_{Edd} = \frac{4\pi G_N M c m_p}{\sigma_T}~~~~.
\end{equation}

Now, consider $\dot{M}_{Edd}$ is the Eddington mass accretion rate of the considered system. If the $\epsilon$ fraction of the mass is supposed to generate energy, then,

\begin{equation}
\label{eqn:edington}
L_{Edd} = \epsilon \dot{M}_{Edd}c^2~~\impliess \dot{M}_{Edd} =  \frac{4\pi G M m_p}{\epsilon c \sigma_T}~~~~.
\end{equation}

To choose the mass $M$ of the central BH, we will enlist a few of them in table~\ref{table:smbh:mass:list}.

\begin{table}[h]
	\centering
	\begin{tabular}{c|c|c}\hline 
		Name & Mass($M_\odot$) & Ref\\ \hline 
		SDSS J102325.31+514251.0  &  $(3.31\pm 0.61)\times10^{10}$  & \cite{wiki13:zuo:2015}\\ \hline
		Messier 87  & $7.22^{+0.34}_{-0.40}\times 10^9$ & \cite{wiki37:10:1093:mnras:stv2982} \\ \hline
		PG 1700+518 & $7.81^{+1.82}_{-1.65}\times 10^8$ & \cite{wiki4:peterson:2014} \\ \hline
		Messier 81 & $7\times 10^7$ & \cite{wiki75:devereux:2003} \\ \hline
		Sagittarius A* & $4.3\times 10^6$ & \cite{wiki88:Ghez:2008}\\ \hline
	\end{tabular}
	\caption{Masses of some SMBHs}
	\label{table:smbh:mass:list}
\end{table}

We generalize this as $M=10^{6+\sigma}\times M_\odot$.

To determine $\epsilon$, we point out the value as $0.01-0.1L_{Edd}$ for quasars and $0.001-0.3$ for Seyfert galaxies. So, we choose
\begin{equation}
\label{eqn:eps}
\epsilon = 3\times 10^{-1-\psi}~~~~.
\end{equation}

Combining equations~(\ref{eqn:edington}) \&~(\ref{eqn:eps}), we have

\begin{equation}
\dot{M}_{Edd} = \frac{4\pi G M_{\odot}m_p 10^7}{3 c \sigma_T} 10^{\sigma+\psi}~~~~.
\end{equation}

Now, let us assume the accretion disc concerned in this work is consuming mass at Eddington mass accretion limit. Then,

\begin{equation}
\rho = \frac{\dot{M}_{Edd}\sqrt{\frac{F_g}{x^3}}}{u\;  c_s}~~\impliess \rho = \frac{4\pi m_p 10^{1+\psi}}{3 c \sigma_T}\times \frac{\sqrt{\frac{F_g}{x^3}}}{u c_s}~~~~.
\end{equation}

\begin{figure}
	\centering
	\setcounter{subfigure}{0}
	\subfigure[$\Gamma=1.6,    \alpha_{ss}=10^{-4}$]{\includegraphics[width=0.24\textwidth]{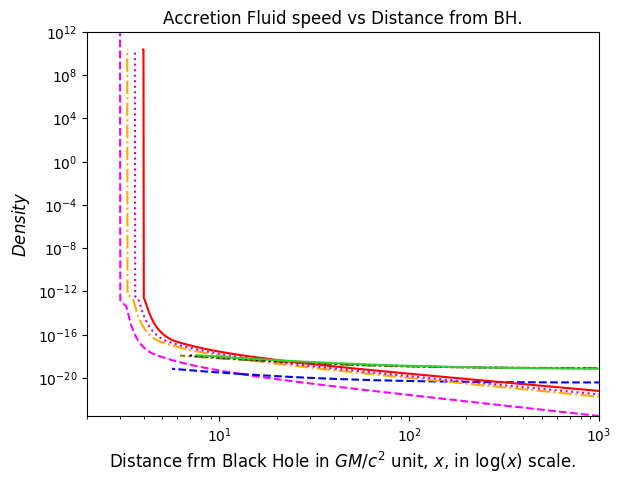}} 
	\subfigure[$\Gamma=1.6,    \alpha_{ss}=10^{-2}$]{\includegraphics[width=0.24\textwidth]{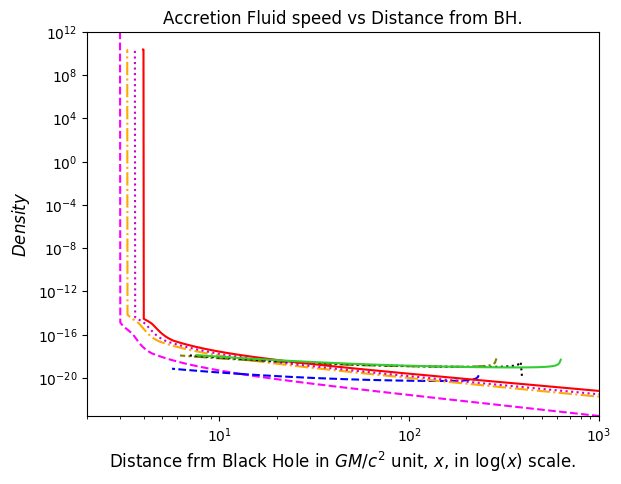}} 
	\subfigure[$\Gamma=0.09,    \alpha_{ss}=10^{-4}$]{\includegraphics[width=0.24\textwidth]{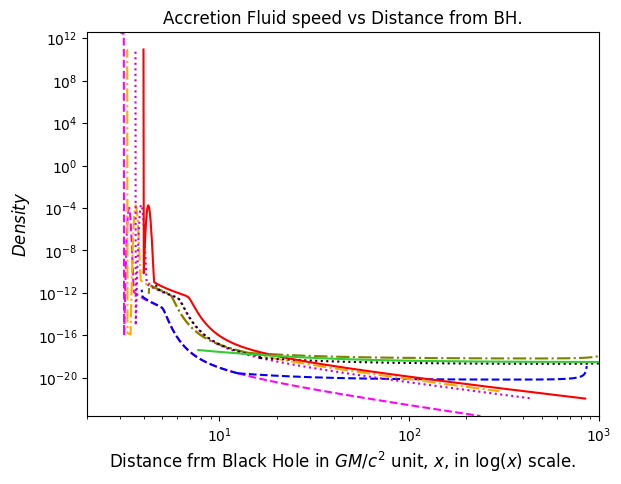}}
	\subfigure[$\Gamma=0.09,    \alpha_{ss}=10^{-2}$]{\includegraphics[width=0.24\textwidth]{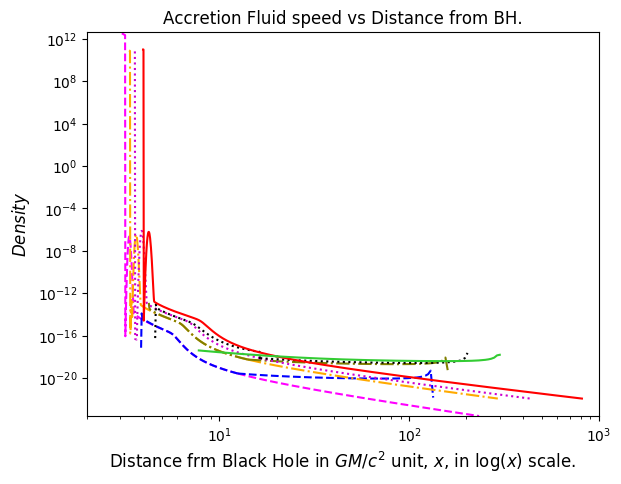}}
	\caption*{\textbf{\emph{Figure 4.1:}} Images for $\lambda_c=2.7, \omega_q = 1/3, A_q=0.01$. Red solid line shows wind for $a=0$,  Green solid line shows accretion for $a=0$, Purple dotted line shows wind for $a=0.5$,  Black dotted line shows accretion for $a=0.5$, Orange dash-dotted line shows wind for $a=0.9$,  Olive dash-dotted line shows accretion for $a=0.9$ and Pink dashed-dashed line shows wind for $a=0.998$,  Blue dashed-dashed line shows accretion for $a=0.998$}
	\setcounter{subfigure}{0}
	\subfigure[$\Gamma=1.6,    \alpha_{ss}=10^{-4}$]{\includegraphics[width=0.24\textwidth]{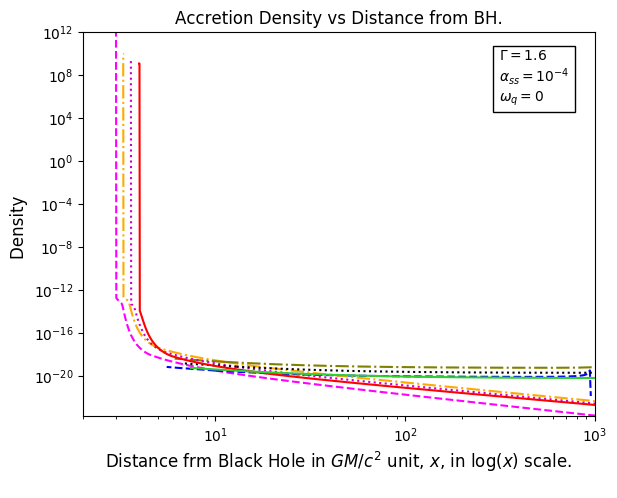}} 
	\subfigure[$\Gamma=1.6,    \alpha_{ss}=10^{-2}$]{\includegraphics[width=0.24\textwidth]{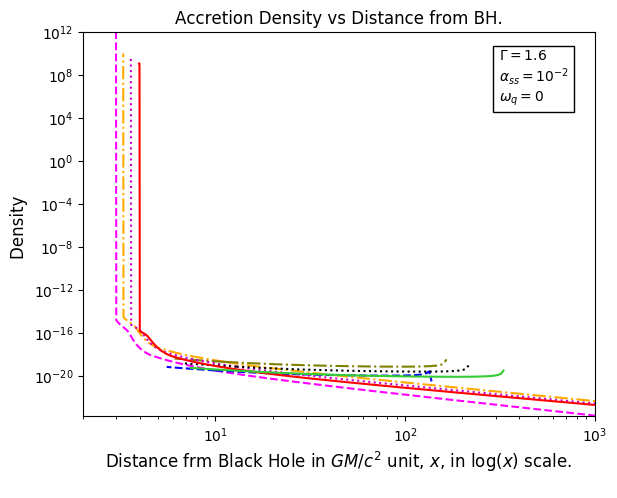}} 
	\subfigure[$\Gamma=0.09,    \alpha_{ss}=10^{-4}$]{\includegraphics[width=0.24\textwidth]{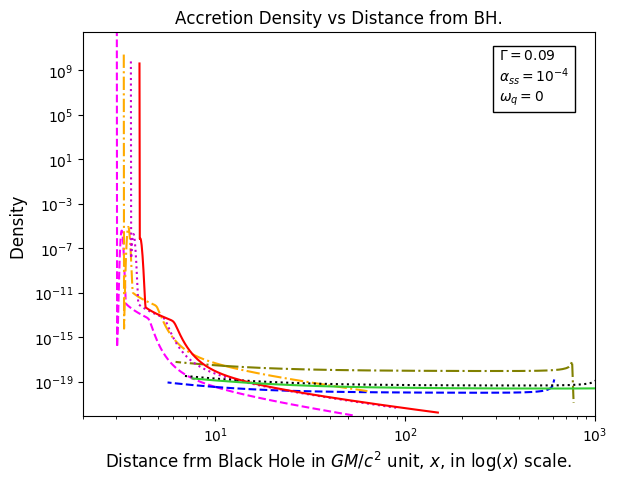}}
	\subfigure[$\Gamma=0.09,    \alpha_{ss}=10^{-2}$]{\includegraphics[width=0.24\textwidth]{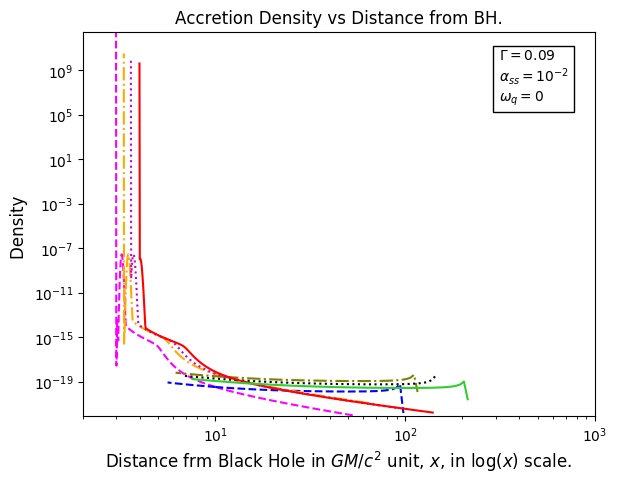}}
	\caption*{\textbf{\emph{Figure 4.2:}} Images for $\lambda_c=2.7, \omega_q = 0, A_q=0.01$. Red solid line shows wind for $a=0$,  Green solid line shows accretion for $a=0$, Purple dotted line shows wind for $a=0.5$,  Black dotted line shows accretion for $a=0.5$, Orange dash-dotted line shows wind for $a=0.9$,  Olive dash-dotted line shows accretion for $a=0.9$ and Pink dashed-dashed line shows wind for $a=0.998$,  Blue dashed-dashed line shows accretion for $a=0.998$}
	\setcounter{subfigure}{0}
	\subfigure[$\Gamma=1.6,    \alpha_{ss}=10^{-4}$]{\includegraphics[width=0.24\textwidth]{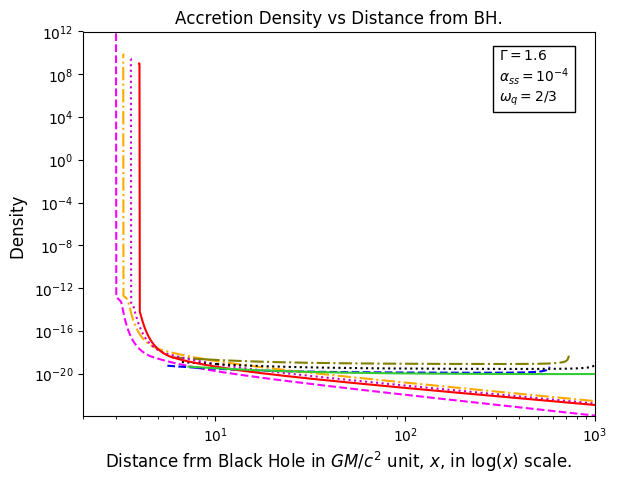}} 
	\subfigure[$\Gamma=1.6,    \alpha_{ss}=10^{-2}$]{\includegraphics[width=0.24\textwidth]{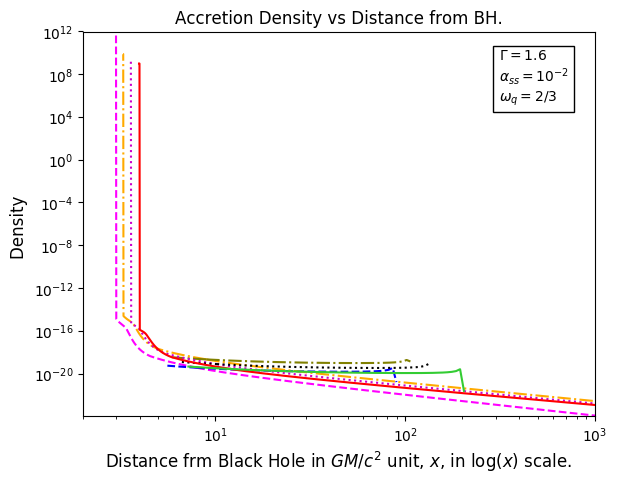}} 
	\subfigure[$\Gamma=0.09,    \alpha_{ss}=10^{-4}$]{\includegraphics[width=0.24\textwidth]{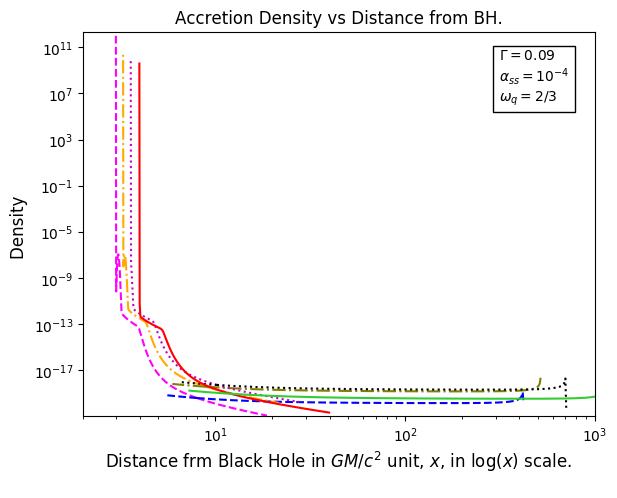}}
	\subfigure[$\Gamma=0.09,    \alpha_{ss}=10^{-2}$]{\includegraphics[width=0.24\textwidth]{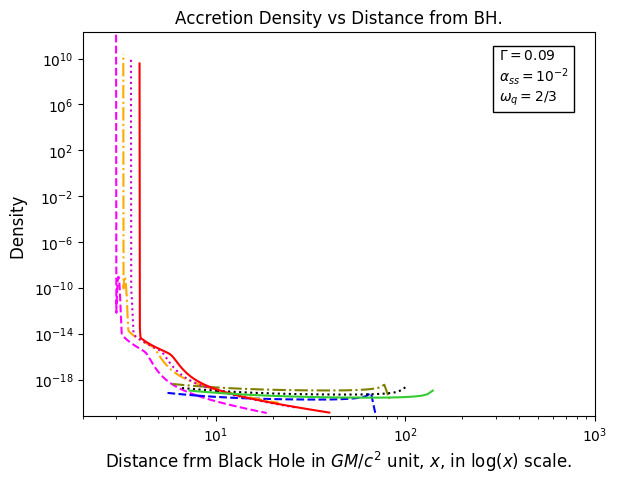}}
	\caption*{\textbf{\emph{Figure 4.3:}} Images for $\lambda_c=2.7, \omega_q = -2/3, A_q=10^{-10}$. Red solid line shows wind for $a=0$,  Green solid line shows accretion for $a=0$, Purple dotted line shows wind for $a=0.5$,  Black dotted line shows accretion for $a=0.5$, Orange dash-dotted line shows wind for $a=0.9$,  Olive dash-dotted line shows accretion for $a=0.9$ and Pink dashed-dashed line shows wind for $a=0.998$,  Blue dashed-dashed line shows accretion for $a=0.998$}
	\caption{Curves for $\eta/s$ vs radial distance from BH.}
\end{figure}

\begin{figure}
	\ContinuedFloat
	\setcounter{subfigure}{0}
	\subfigure[$\Gamma=1.6,    \alpha_{ss}=10^{-4}$]{\includegraphics[width=0.24\textwidth]{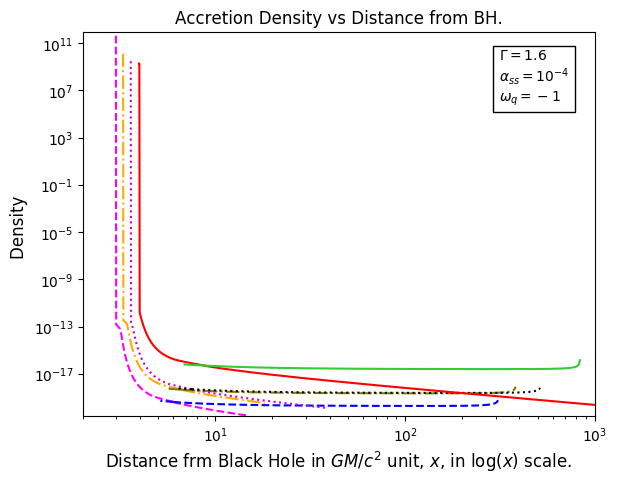}} 
	\subfigure[$\Gamma=1.6,    \alpha_{ss}=10^{-2}$]{\includegraphics[width=0.24\textwidth]{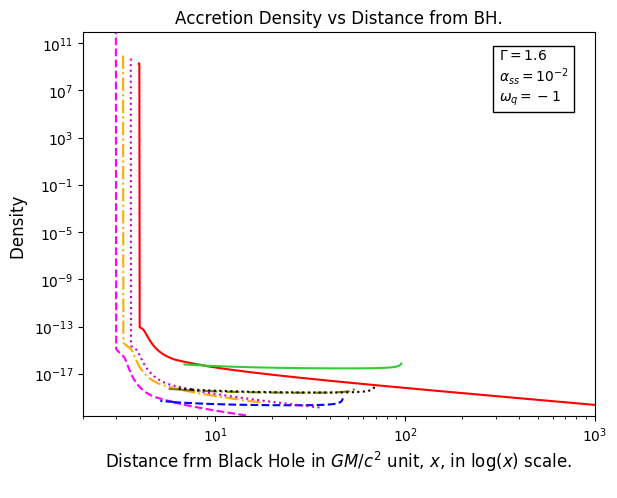}} 
	\subfigure[$\Gamma=0.09,    \alpha_{ss}=10^{-4}$]{\includegraphics[width=0.24\textwidth]{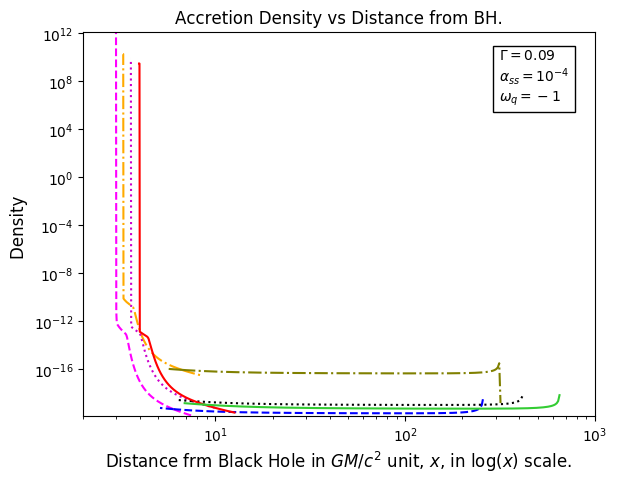}}
	\subfigure[$\Gamma=0.09,    \alpha_{ss}=10^{-2}$]{\includegraphics[width=0.24\textwidth]{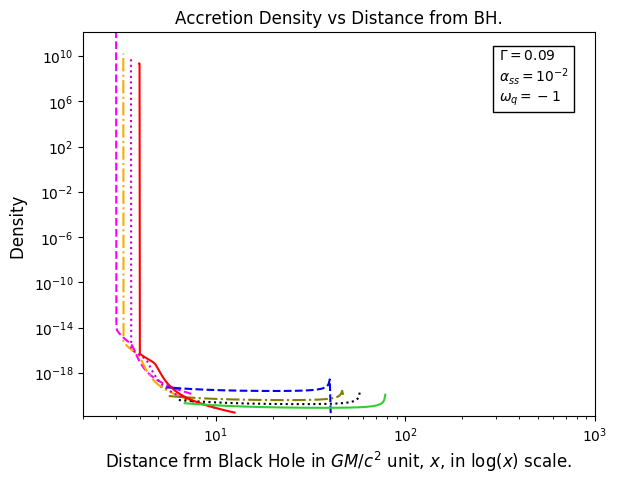}}
	\caption*{\textbf{\emph{Figure 4.4:}} Images for $\lambda_c=2.7, \omega_q = -1, A_q=10^{-10}$. Red solid line shows wind for $a=0$,  Green solid line shows accretion for $a=0$, Purple dotted line shows wind for $a=0.5$,  Black dotted line shows accretion for $a=0.5$, Orange dash-dotted line shows wind for $a=0.9$,  Olive dash-dotted line shows accretion for $a=0.9$ and Pink dashed-dashed line shows wind for $a=0.998$,  Blue dashed-dashed line shows accretion for $a=0.998$}
	\caption{Curves for $\eta/s$ vs radial distance from BH.}
	\label{fig:density}
\end{figure}

\begin{figure}
	\centering
	\setcounter{subfigure}{0}
	\subfigure[$\Gamma=1.6,    \alpha_{ss}=10^{-4}$]{\includegraphics[width=0.24\textwidth]{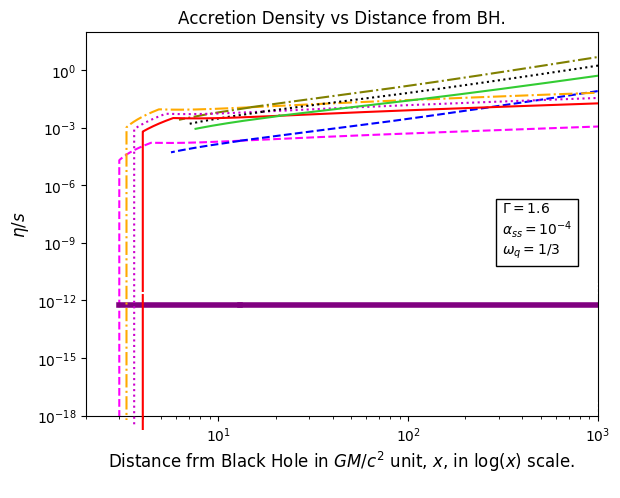}} 
	\subfigure[$\Gamma=1.6,    \alpha_{ss}=10^{-2}$]{\includegraphics[width=0.24\textwidth]{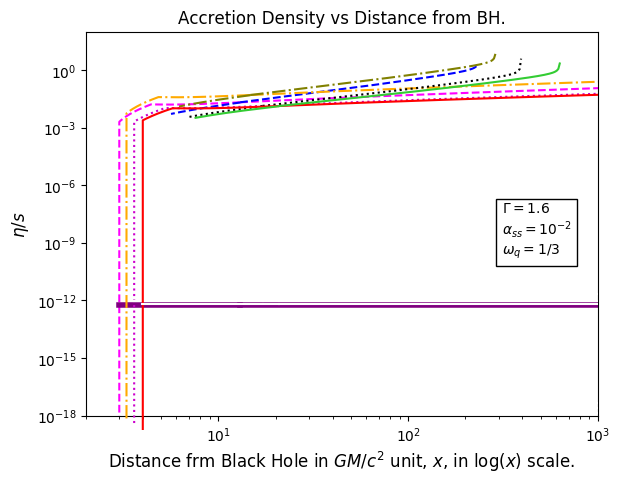}} 
	\subfigure[$\Gamma=0.09,    \alpha_{ss}=10^{-4}$]{\includegraphics[width=0.24\textwidth]{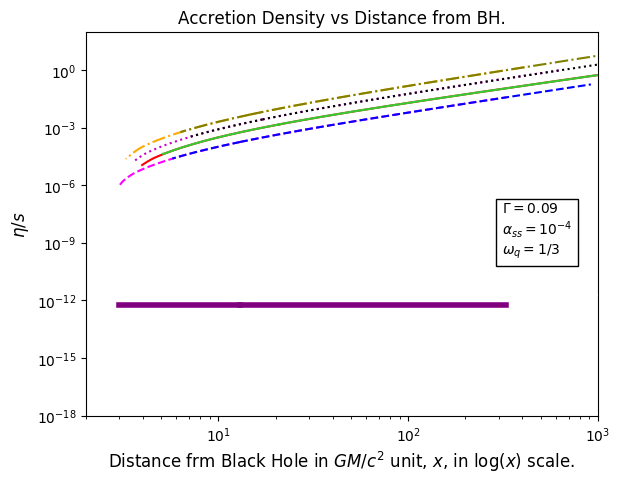}}
	\subfigure[$\Gamma=0.09,    \alpha_{ss}=10^{-2}$]{\includegraphics[width=0.24\textwidth]{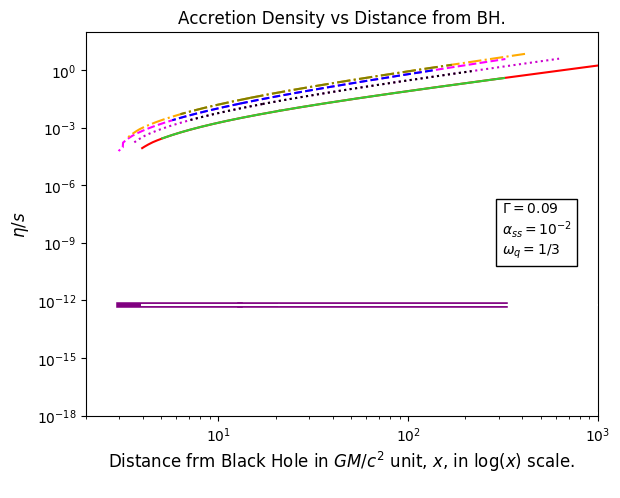}}
	\caption*{\textbf{\emph{Figure 5.1:}} Images for $\lambda_c=2.7, \omega_q = 1/3, A_q=0.01$. Red solid line shows wind for $a=0$,  Green solid line shows accretion for $a=0$, Purple dotted line shows wind for $a=0.5$,  Black dotted line shows accretion for $a=0.5$, Orange dash-dotted line shows wind for $a=0.9$,  Olive dash-dotted line shows accretion for $a=0.9$ and Pink dashed-dashed line shows wind for $a=0.998$,  Blue dashed-dashed line shows accretion for $a=0.998$}
	\setcounter{subfigure}{0}
	\subfigure[$\Gamma=1.6,    \alpha_{ss}=10^{-4}$]{\includegraphics[width=0.24\textwidth]{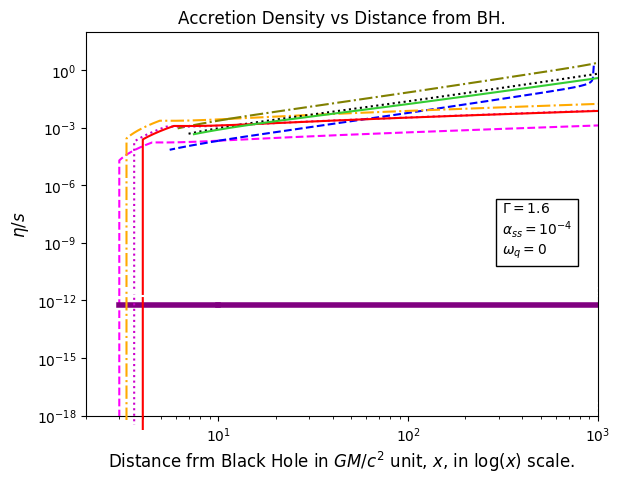}} 
	\subfigure[$\Gamma=1.6,    \alpha_{ss}=10^{-2}$]{\includegraphics[width=0.24\textwidth]{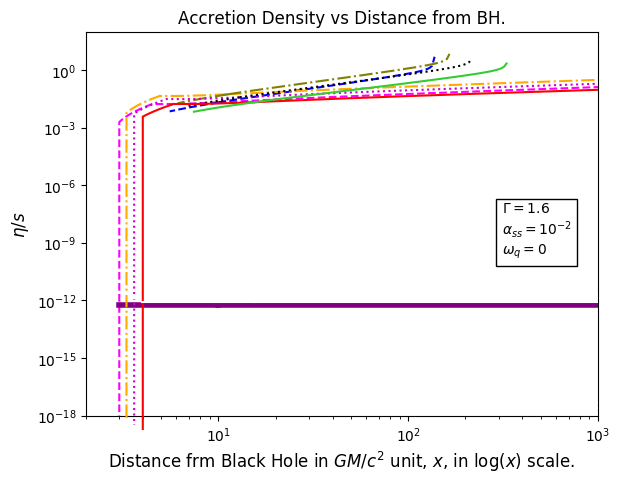}} 
	\subfigure[$\Gamma=0.09,    \alpha_{ss}=10^{-4}$]{\includegraphics[width=0.24\textwidth]{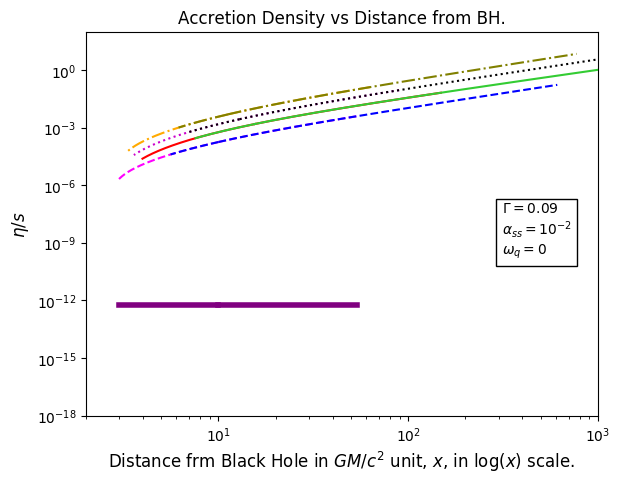}}
	\subfigure[$\Gamma=0.09,    \alpha_{ss}=10^{-2}$]{\includegraphics[width=0.24\textwidth]{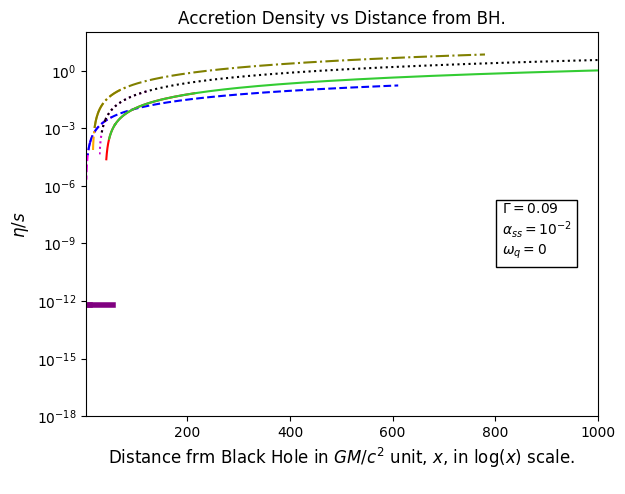}}
	\caption*{\textbf{\emph{Figure 5.2:}} Images for $\lambda_c=2.7, \omega_q = 0, A_q=0.01$. Red solid line shows wind for $a=0$,  Green solid line shows accretion for $a=0$, Purple dotted line shows wind for $a=0.5$,  Black dotted line shows accretion for $a=0.5$, Orange dash-dotted line shows wind for $a=0.9$,  Olive dash-dotted line shows accretion for $a=0.9$ and Pink dashed-dashed line shows wind for $a=0.998$,  Blue dashed-dashed line shows accretion for $a=0.998$}
	\setcounter{subfigure}{0}
	\subfigure[$\Gamma=1.6,    \alpha_{ss}=10^{-4}$]{\includegraphics[width=0.24\textwidth]{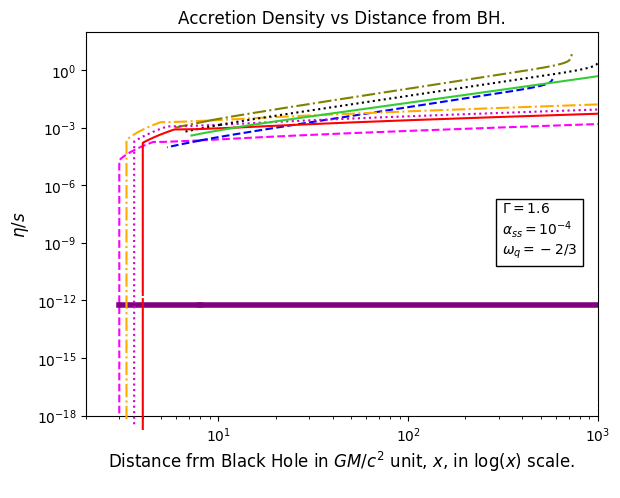}} 
	\subfigure[$\Gamma=1.6,    \alpha_{ss}=10^{-2}$]{\includegraphics[width=0.24\textwidth]{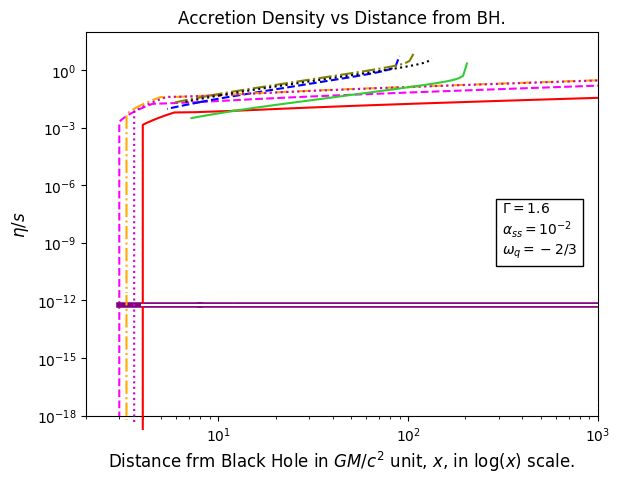}} 
	\subfigure[$\Gamma=0.09,    \alpha_{ss}=10^{-4}$]{\includegraphics[width=0.24\textwidth]{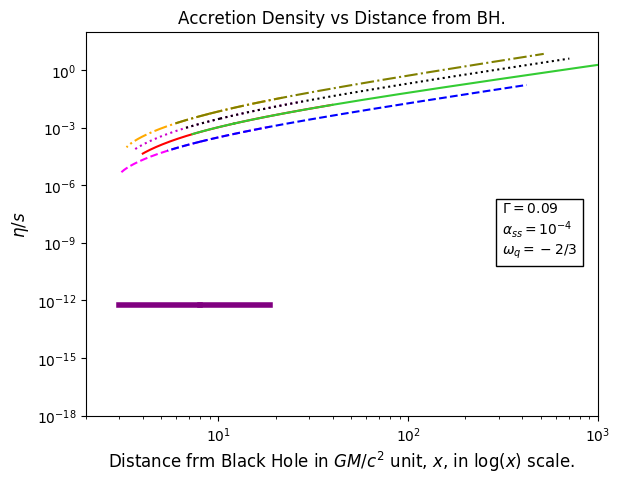}}
	\subfigure[$\Gamma=0.09,    \alpha_{ss}=10^{-2}$]{\includegraphics[width=0.24\textwidth]{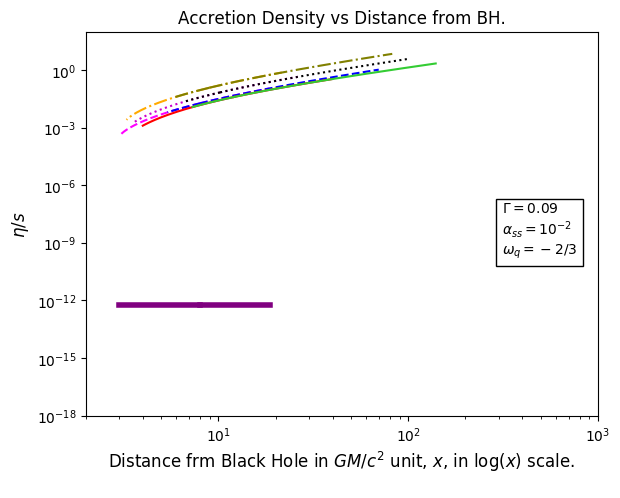}}
	\caption*{\textbf{\emph{Figure 5.3:}} Images for $\lambda_c=2.7, \omega_q = -2/3, A_q=10^{-10}$. Red solid line shows wind for $a=0$,  Green solid line shows accretion for $a=0$, Purple dotted line shows wind for $a=0.5$,  Black dotted line shows accretion for $a=0.5$, Orange dash-dotted line shows wind for $a=0.9$,  Olive dash-dotted line shows accretion for $a=0.9$ and Pink dashed-dashed line shows wind for $a=0.998$,  Blue dashed-dashed line shows accretion for $a=0.998$}
	\caption{Curves for $\eta/s$ vs radial distance from BH.}
\end{figure}

\begin{figure}
	\ContinuedFloat
	\setcounter{subfigure}{0}
	\subfigure[$\Gamma=1.6,    \alpha_{ss}=10^{-4}$]{\includegraphics[width=0.24\textwidth]{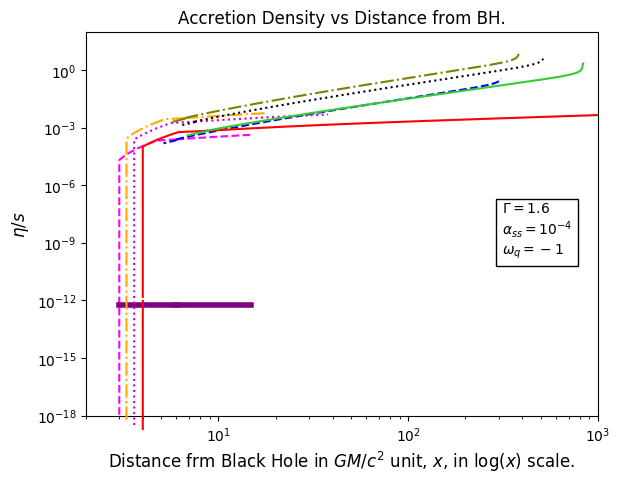}} 
	\subfigure[$\Gamma=1.6,    \alpha_{ss}=10^{-2}$]{\includegraphics[width=0.24\textwidth]{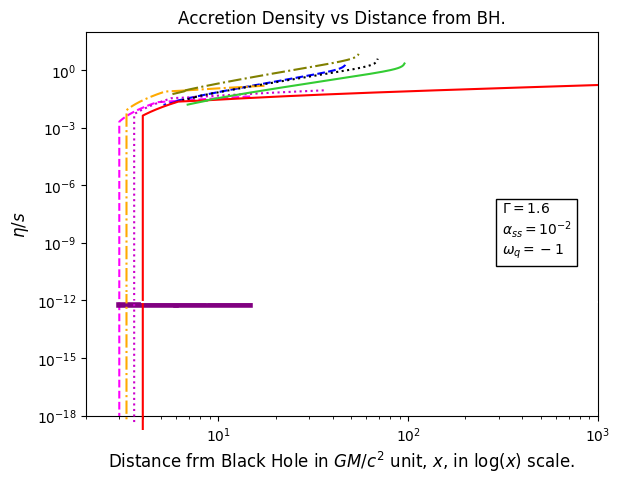}} 
	\subfigure[$\Gamma=0.09,    \alpha_{ss}=10^{-4}$]{\includegraphics[width=0.24\textwidth]{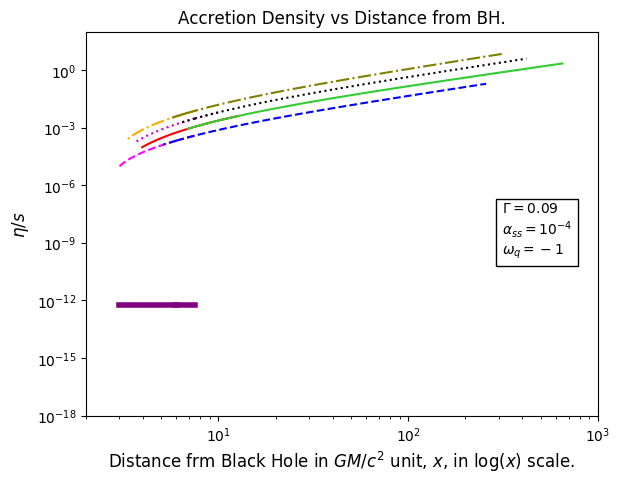}}
	\subfigure[$\Gamma=0.09,    \alpha_{ss}=10^{-2}$]{\includegraphics[width=0.24\textwidth]{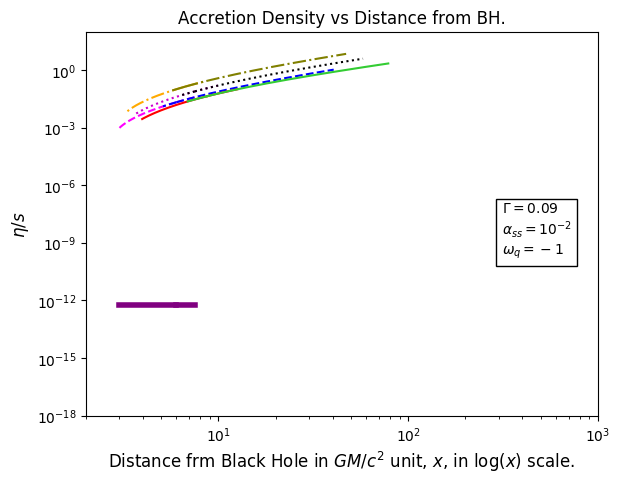}}
	\caption*{\textbf{\emph{Figure 5.4:}} Images for $\lambda_c=2.7, \omega_q = -1, A_q=10^{-10}$. Red solid line shows wind for $a=0$,  Green solid line shows accretion for $a=0$, Purple dotted line shows wind for $a=0.5$,  Black dotted line shows accretion for $a=0.5$, Orange dash-dotted line shows wind for $a=0.9$,  Olive dash-dotted line shows accretion for $a=0.9$ and Pink dashed-dashed line shows wind for $a=0.998$,  Blue dashed-dashed line shows accretion for $a=0.998$}
	\caption{Curves for $\eta/s$ vs radial distance from BH.}
	\label{fig:eta}
\end{figure}

We plot density vs radial distances in figures 4.1.a to 4.4.d. Among them, figures 4.1.a to 4.1.d are for the EoS value $\omega_q=\frac{1}{3}$. In figure 4.1.a, where viscosity is low ($\alpha_{SS}=10^{-4}$), the wind density varies from $10^{-21}$ to $10^{11}gmcm^{-3}$. In figure 4.1.b, viscosity is high ($\alpha_{SS}=10^{-2}$) and as a result we observe the wind density to raise from the order of $10^{-24}gmcm^{-3}$ at $10^{3}$ Schwarzschild radius to $10^{12}gmcm^{-3}$ near the event horizon. For low value of $\Gamma$, we observe that the wind varies for a larger range.

Accretion density, however, varies for a comparatively smaller range.

The same wide range of variations are observed for all the $\omega_q$ cases.

We can explain this issue like : the accretion density varies through an order of $10^{4}$ centering commonly around the value $10^{-18}gmcm^{-3}$. The negative pressure of quintessence may oppose the accreting matter to fall in and hence near the BH, the wind speed is tremendously high. Wind density profiles which does match with the density profile predictions of the article like~\cite{di:matteo:2003}.

\subsection{Shear Viscosity Coefficient to Entropy Density Ratio}
\label{subsec:viscosity}
Dual holographic nature of states is predicted by strongly interacting quantum field theories. For example, we can choose the systems where BHs are embedded in AdS space. For such a system, a universal lower bound of the shear viscosity coefficient ($\eta$) to entropy density ($s$) ratio is prescribed as~\cite{kovtun:2005},\cite{policastro:2001},\cite{tamaryan:2003},\cite{buchel:2004},\cite{son:2007} $$\frac{\eta}{s}\geq \frac{1}{4\pi} \frac{\hbar}{\kappa_B}$$
This lower bound is popularly known as the Kovtun-Starinets-Son (KSS) bound.

However, Jakovac~\cite{jakovac:2010} has calculated the $\frac{\eta}{s}$ ratio mathematically. To do this, he has assumed some physical conditions for the spectral functions and kept the entropy density constant. He observed that the lower bound may not be universal for some systems which carry quasi-particle constituents with small wave function re-normalization constant, high temperature strongly interacting systems or systems with low temperatures and zero mass excitation. As we have discussed in the introductory section of this article, DE may possess a very small amount of shear viscosity. Besides, the entropy density for DE should be high due to its repulsive nature. Entropy for different components of universe shows~\cite{egan:2010}. That for cosmic  event horizon it may raise up to $2.6\pm 0.3\times 10^{122}$ and for SMBHs it may go up to $1.2^{+1.1}_{-0.7}\times 10^{103}$. So for a phenomenon which involves both SMBHs and DE, the $\frac{\eta}{s}$ ratio may fall and can become lower than the theoretical prediction as well.

We follow the reference~\cite{151:banibrata:2013151} to set the entropy equation as
\begin{equation}
uT\frac{ds}{dx}=q_{vis}+q_{mag}+q_{nex}-q_{rad}~~~~.
\end{equation}

$T$ denotes the temperature of the flow. $s$ is the entropy density. $q_{vis}$, $q_{mag}$ and $q_{nex}$ respectively denote the energies released per unit volume per unit time due to viscous dissipation, magnetic dissipation and thermonuclear reactions. $q_{rad}$ indicates the energy radiated away per unit volume per unit time by various cooling process like synchrotron, bremsstrahlung and inverse Comptonisation of soft photons and the energy absorbed per unit volume per unit time due to thermonuclear reactions. 

As a result, the entropy density of the flow can be expressed as
\begin{equation}
s=\int \frac{q_{vis}+q_{mag}+q_{nex}-q_{rad}}{uT}dx~~~~.
\end{equation}

The turbulent kinematic viscosity $\gamma$ can be scaled linearly with sound speed, $c_s$, in the flow and half thickness, $h$, of the disc, providing $\gamma=\alpha_{SS}c_s h$. Hence the shear viscosity has the form 

\begin{equation}
\gamma=\alpha_{SS}\rho c_s h
\end{equation}

and ultimately, the $\frac{\eta}{s}$ ratio for DE accretion is found to be

\begin{equation}
\frac{\eta}{s}=\frac{\alpha_{SS}c_s^2\sqrt{\frac{x_c}{F_g}}\rho^{n+1}}{\rho^{n+1}\left(\alpha+1\right)-\beta}~~~~.
\end{equation}

\section{Brief Discussions and Conclusions}
\label{sec:conclusion}
This present article can be treated as a detailed study of the viscous accretion onto a rotating black hole embedded in a quintessence universe and the consequent thermodynamic phenomena. To construct the mathematical model we have chosen a particular type of black hole which has mass and rotation as signature properties along with a special type of background. Quintessence is a hypothetical fluid which is theorized to create repulsive force responsible for late time cosmic acceleration. We choose a rotating black hole solution which carries effects of quintessence universe in it. The gravitational effect of such a black hole is implied through a pseudo Newtonian potential. This is done as direct general relativistic nonlinear differential equations are difficult to solve. Viscous effect is adopted through the Shakura and Sunyaev $\alpha_{SS}$ effects.

We follow that if the viscosity is high the accretion branch's fluid speed steeply falls down as we go far from the central black hole. Wind speed increases as we increase viscosity. But the radial distance wise shift is small. At a finite distance fluid speed becomes equal to that of light. As we increase the quintessential effect, wind speed increases. 

Truncation in the accretion length is supported by the sonic speed curves and specific angular momentum to Keplerian angular momentum ratio curves. Either the sonic speed reaches the speed of light or the $\frac{\lambda}{\lambda_k}$ ratio reaches the value $1$ where the accretion turns zero. Steep fall in accretion due to the increase in viscosity signifies the weakening of accretion procedure.

Density profiles are found to be very interesting. At the edges of the disc, approximately at the order of thousand Schwarzschild radius distance the density is found to be very low. But of course this was higher than the density of universe. At the nearer vicinity of the SMBH, we see the wind density to raise up to the order of $10^{12}gmcm^{-3}$. This quite matches with the observational results. 

Finally, we study the $\frac{\eta}{s}$ ratio and follow that this ratio turns to be less than the theoretical predictions. Present day speculation about the shear viscosity of DE supports this result. Interestingly, we achieve the result where $\eta/s$ is lower than the pre-predicted value for adiabatic accretion. Only the dark energy contamination is considered for the BH metric itself, not onto the accreting fluid's property. Our result strongly states that far late time BHs, accretion of adiabatic fluid can even reduce the $\eta/s$ ratio.

\section*{Acknowledgment}

This research is supported by the project grant of Government of West Bengal, Department of Higher Education, Science and Technology and Biotechnology (File no:- $ST/P/S\&T/16G-19/2017$). RB thanks IUCAA for providing Visiting Associateship.

\bibliographystyle{unsrt}


\end{document}